\documentclass[a4paper,fleqn,usenatbib]{mnras}

\usepackage{newtxtext,newtxmath}

\usepackage[T1]{fontenc}
\usepackage{ae,aecompl}

\usepackage{graphicx}	
\usepackage{amsmath}	
\usepackage{amssymb}	

\usepackage{gensymb}    
\usepackage{booktabs}   
\usepackage{url}   
\usepackage{cleveref}
\usepackage{subfig}
\usepackage{threeparttable}
\usepackage{lipsum}
\usepackage[percent]{overpic}
\usepackage{pict2e}
\usepackage{fixltx2e}
\usepackage{xspace}
\usepackage[usenames,dvipsnames]{xcolor}

\newcommand{\hst}{\textit{HST}\xspace}
\newcommand{\mdot}{\mbox{$\dot{M}$}}
\newcommand{\msun}{\mbox{$\mathrm{M}_\odot$}}
\newcommand{\mwd}{\mbox{$M_\mathrm{WD}$}}
\newcommand{\rsun}{\mbox{$\mathrm{R}_\odot$}}
\newcommand{\rwd}{\mbox{$R_\mathrm{WD}$}}
\newcommand{\porb}{\mbox{$P_\mathrm{orb}$}}
\newcommand{\twd}{\mbox{$T_\mathrm{WD}$}}
\newcommand{\kms}{\mbox{$\mathrm{km\,s}^{-1}$}}

\usepackage{soul}

\title[Spiral shocks in the accretion disc of SDSS1238]{\vspace{-0.8cm}
Evidence for mass accretion driven by spiral shocks onto the white dwarf in SDSS\,J123813.73--033933.0}

\author[A.~F.~Pala et al.]{A.~F.~Pala,$^{1,2}$\thanks{E-mail: apala@eso.org}
B.~T.~G\"ansicke,$^{2}$
T.~R.~Marsh,$^{2}$
E.~Breedt,$^{3}$ 
J.~J.~Hermes,$^{4}$\thanks{Hubble Fellow} \newauthor
J.~D.~Landstreet,$^{5,6}$ 
M.~R.~Schreiber,$^{7}$
D.~M.~Townsley,$^{8}$
L.~Wang,$^{2}$ 
A.~Aungwerojwit,$^{9}$\newauthor
F.--J.~Hambsch,$^{10,11,12}$ 
B.~Monard,$^{13}$ 
G.~Myers,$^{12}$ 
P.~Nelson,$^{12}$
R.~Pickard,$^{14}$ 
G.~Poyner,$^{14}$\newauthor
D.~E.~Reichart,$^{4}$
R.~Stubbings,$^{12}$
P.~Godon,$^{15,16}$
P.~Szkody,$^{17}$
D.~De~Martino,$^{18}$\newauthor
V.~S.~Dhillon,$^{19,20}$
C.~Knigge,$^{21}$
S.~G.~Parsons$^{19}$\\
$^{1}$European Southern Observatory, Karl Schwarzschild Stra{\ss}e 2, Garching, 85748, Germany\\
$^{2}$Department of Physics, University of Warwick, Coventry, CV4 7AL, UK\\
$^{3}$Institute of Astronomy, University of Cambridge, Cambridge, CB3 0HA, UK\\
$^{4}$Department of Physics and Astronomy, University of North Carolina, Chapel Hill, NC 27599, USA\\
$^{5}$Armagh Observatory, College Hill, Armagh, BT61 9DG, UK\\
$^{6}$Department of Physics \& Astronomy, University of Western Ontario, London, Ontario, N6A 3K7, Canada\\
$^{7}$Instituto de F{\'i}sica y Astronom{\'i}a, Universidad de Valpara{\'i}so, 2360102 Valparaiso, Chile\\
$^{8}$Department of Physics and Astronomy, University of Alabama, Tuscaloosa, AL 35405, USA\\
$^{9}$Department of Physics, Faculty of Science, Naresuan University, Phitsanulok 65000, Thailand\\
$^{10}$Vereniging Voor Sterrenkunde (VVS), Oude Bleken 12, 2400 Mol, Belgium\\
$^{11}$Bundesdeutsche Arbeitsgemeinschaft f{\"u}r Ver{\"a}nderliche Sterne (BAV), Munsterdamm 90, 12169 Berlin, Germany\\
$^{12}$American Association of Variable Star Observers, Cambridge, MA 02138, USA\\
$^{13}$CBA Kleinkaroo, Calitzdorp, South Africa\\
$^{14}$British Astronomical Association, Variable Star Section, Burlington House, Piccadilly, London, W1J ODU, UK\\
$^{15}$Astronomy \& Astrophysics, Villanova University, Villanova, PA 19085, USA\\
$^{16}$Rowland Department of Physics \& Astronomy, The Johns Hopkins University, Baltimore, MD 21218, USA\\
$^{17}$Department of Astronomy, University of Washington, Seattle, WA 98195--1580, USA\\
$^{18}$INAF -- Osservatorio Astronomico di Capodimonte, Napoli, I--80131, Italy\\
$^{19}$Department of Physics and Astronomy, University of Sheffield, Sheffield S3 7RH, UK\\
$^{20}$Instituto de Astrofisica de Canarias, E-38205 La Laguna, Tenerife, Spain\\
$^{21}$School of Physics and Astronomy, University of Southampton, Southampton, SO17 1BJ, UK\\}
\date{\vspace*{-0.5cm} Accepted 2018 November 14. Received 2018 November 14; in original form 2018 March 24.}
\pubyear{2018}

\begin{document}
\label{firstpage}
\pagerange{\pageref{firstpage}--\pageref{lastpage}}
\maketitle

\vspace*{-1.8cm} {
\begin{abstract}
We present high-time--resolution photometry and phase--resolved spectroscopy of the short-period ($\porb = 80.52\,\mathrm{min}$) cataclysmic variable SDSS\,J123813.73--033933.0, observed with the \textit{Hubble Space Telescope} (\hst), the \textit{Kepler}/\textit{K2} mission and the Very Large Telescope (VLT). We also report observations of the first detected super-outburst. 
SDSS\,J1238--0339 shows two types of variability: quasi-regular brightenings recurring every $\simeq8.5$\,h during which the system increases in brightness by $\simeq 0.5\,$mag, and a double hump quasi-sinusoidal modulation at the orbital period. The detailed \textit{K2} light curve reveals that the amplitude of the double-humps increases during the brightenings and that their phase undergoes a $\simeq 90^{\circ}$ phase shift with respect to the quiescent intervals. The \hst data unambiguously demonstrate that these phenomena both arise from the heating and cooling of two relatively large regions on the white dwarf. We suggest that the double-hump modulation is related to spiral shocks in the accretion disc resulting in an enhanced accretion rate heating two localised regions on the white dwarf, with the structure of the shocks fixed in the binary frame explaining the period of the double humps. The physical origin of the 8.5\,h brightenings is less clear. However, the correlation between the observed variations of the amplitude and phase of the double-humps with the occurrence of the brightenings is supportive of an origin in thermal instabilities in the accretion disc.
\end{abstract}}

\pagebreak

\section{Introduction}
Cataclysmic variables (CVs) are close interacting binaries containing a white dwarf, the primary star, and a low--mass companion, the donor or secondary star \citep{Warner}. In these systems, the secondary star fills its Roche lobe and loses mass through the inner Lagrangian point. If the white dwarf is not strongly magnetic ($B \lesssim 10\,$MG), accretion onto the white dwarf proceeds via an accretion disc. 

The secular evolution of CVs is driven by angular momentum loss which, by continuously shrinking the system, keeps the secondary in touch with its Roche lobe, sets the mass transfer rate ($\mdot$) onto the white dwarf and results in a decrease of the orbital period ($\porb$, \citealt{Rappaport1983,Paczynski_Sienkiewicz_1983,SpruitRitter}). Hence, CVs evolve from long to short orbital periods, until they reach the ``period minimum'' \citep{Rappaport_Joss_Webbink_1982,Paczynski_Sienkiewicz_1983}. Two timescales determine the occurrence of this minimum period: the rate at which the mass is stripped of the secondary (mass--loss timescale, $\tau_{\dot{M}_2}$) and the time required for the donor to adjust its radius corresponding to its decreasing mass (the thermal timescale, $\tau_{\mathrm{kh}}$). When $\tau_{\dot{M}_2}$ is much longer than $\tau_{\mathrm{kh}}$, the secondary has the time to find its thermal equilibrium and the system evolves from long to short orbital periods. If, in contrast, $\tau_{\dot{M}_2} \ll\tau_{\mathrm{kh}}$, the secondary is driven out of thermal equilibrium and stops shrinking in response to mass loss. Consequently the system evolves back towards longer orbital periods, becoming a ``period bouncer''. Binary population synthesis studies predict the period minimum to occur at $\simeq 70\,\mathrm{min}$ \citep{Paczynski_1981,Paczynski_Sienkiewicz_1981,Kolb_baraffe_1999,Howell_et_al_2001,Knigge2011,Goliasch-Nelson2015,Kalomeni2016}. Systems at this orbital period are characterised by extreme mass ratios $q = M_2/M_1 \le 0.1$ (where $M_2$ and $M_1$ are the secondary and the white dwarf mass, respectively) and secondary stars with masses of $M_2 \lesssim 0.06-0.07\,\mathrm{M}_\odot$ \citep{Howell_et_al_1997,Knigge2011}.

Over the last decade, the Sloan Digital Sky Survey (SDSS, \citealt{York_et_al_2000}) has led to the spectroscopic identification of over 300 CVs \citep{Szkody_etal_2002, Szkody_etal_2003, Szkody_etal_2004, Szkody_etal_2005, Szkody2006, Szkody_etal_2007, Szkody_etal_2009, Szkody_etal_2011}, many of which are located at the period minimum which \citet{Boris_2009} demonstrated to be in the range $80\,\mathrm{min} \lesssim P_\mathrm{min} \lesssim 86$\,min. Among the CVs discovered by SDSS, SDSS\,J123813.73--033933.0 (aka V406\,Vir, hereafter SDSS1238) stood out because of its particular optical variability. SDSS1238 was identified as a CV in 2003 by \citet{Szkody_etal_2003}, who first estimated its orbital period from seven time--resolved spectra, finding $\porb \simeq 76\,\mathrm{min}$. This value was later refined to $\porb = 80.5 \pm 0.5\,\mathrm{min}$ from a radial velocity study of the H$\alpha$ emission line by \citet{Zharikov_2006}, locating SDSS1238 at the period minimum. The short orbital period, the absence of normal disc outbursts and its recent superoutburst (31st of July 2017, VSNET--alert~21308), make it a member of the WZ\,Sge CV subclass. 

The peculiarity of SDSS1238 resides in its light curve, in which \citet{Zharikov_2006} discovered two types of variability: the first is a sudden increase in brightness, up to $\simeq 0.45\,\mathrm{mag}$, occurring quasi-periodically every 8--12 hours (called ``brightenings'').  The second is a double--humped variation at the orbital period. \citet{Aviles2010} suggested that these double--humps are caused by spiral arms in the accretion disc, which should arise in the disc extending permanently out to the 2:1 resonance radius in systems with a mass ratio $q \le 0.1$ \citep{Lin_Papaloizou_1979}. Similar double--humped light curves have also been observed in other WZ\,Sge CVs, such as V455\,And \citep{Araujo-Betancor_et_al_2005}, GW\,Lib \citep{Woudt_Warner_2002}, SDSS\,J080434.20+510349.2 (hereafter SDSS0804, \citealt{Szkody2006,Zharikov2008}) and WZ\,Sge itself \citep{pattersonetal98-2}. Moreover, GW\,Lib and SDSS0804 also share with SDSS1238 the occurrence of similar brightening events \citep{Woudt_Warner_2002,Zharikov2008,Odette,Chote_Sullivan_2016}, which have also been detected in GP\,Com \citep{Marsh_et_al_1995}.

We present new photometric and spectroscopic observations of SDSS1238 obtained during quiescence, when the white dwarf is accreting material from the disc at a very low rate, using the \textit{Hubble Space Telescope} (\textit{HST}), the Very Large Telescope (VLT) and the \textit{Kepler}/$K2$ mission.  We also report an analysis of the first superoutburst of SDSS1238 and its subsequent return to quiescence, followed-up eight months later with ULTRACAM on the New Technology Telescope (NTT). By investigating the system variability over a wide range of wavelengths (from the far--ultraviolet into the near--infrared) and with high--time resolution photometry, we reveal a fast heating and cooling of a fraction of the white dwarf as the origin of the peculiar light curve of SDSS1238. We discuss these results in the framework of accretion modulation through spiral density shocks and thermal instabilities in the accretion disc.

\section{Observations}\label{sec:obs}
\begin{table*}
  \caption{Summary of the ground--based CCD observations for the monitoring of SDSS1238. }\label{table:obs_log_superoutburst}
 \setlength{\tabcolsep}{0.1cm}
  \begin{tabular}{@{}lccccc@{}}
  \toprule
Telescope                                                     & Observation date               &  Location & Diameter & Filter & Exposure  \\
                                                                    &                                         &                &  (cm)      &         & time (s)    \\ 
\midrule 
Remote Observatory Atacama Desert (ROAD)  & 20 Feb - 13 Mar 2014       & Chile        & 40   & --       & 120     \\
Remote Observatory Atacama Desert (ROAD)  &  4-7 Aug 2017                & Chile        & 50   & --       & 30     \\
Kleinkaroo Observatory                                   &  5-9, 13, 17 Aug 2017    & South Africa & 30   & --        & 30    \\
Prompt 8                                                       &  5-8,  14 - 15 Aug 2017  & Chile        & 61   & $V$    & 15--30 \\
Ellinbank Observatory                                     &  5 Aug 2017                      & Australia    & 31.7 & $V$     & 60 \\
Tetoora Road Observatory                              & 5, 13, 14 Aug 2017           & Australia    & 55   & Visual  &  --  \\
Siding Spring                                                  & 6-8, 13 Aug 2017            & Australia    & 43.1 & $V$    &  30 \\
Las Cumbres Observatory Global Telescope (LCOGT)   & 5, 8 Aug 2017        & Australia    & 40   & $V$     & 60\\            ULTRACAM @ NTT & 14-15 Apr 2018     & Chile          & 358 & Super-$u$, Super-$g$, Super-$r$ & 30,10,10 \\                   
\bottomrule
\end{tabular}
\end{table*}

\begin{table*}
  \caption{Log of the spectroscopic observations.}\label{table:obs_log}
  \begin{tabular}{@{}ccccccc@{}}
  \toprule
  Instrument   & Channel  &  Observation &  Exposure        & Wavelength   &  Slit width/   & Resolution \\
                     &   /         &  date            &   time              &  Range          &  aperture      &  (\AA)     \\ 
                     &  Arm      &                    &    (s)               & (\AA)           &  (\arcsec)     &           \\
  \midrule 
\hst/COS       &  FUV     & 2013 Mar 01  &   7183           & 1150--1730   &   2.5   &  1.08      \\
  \midrule   
               & UVB      & 2015 May 12  &  12 $\times$ 480 & 3000--5595   &   1.0   &  1.1       \\
VLT/X--shooter & VIS      & 2015 May 12  &  10 $\times$ 587 & 5595--10\,240  &   0.9   &  0.8       \\
               & NIR      & 2015 May 12  &  12 $\times$ 520 & 10\,240--24\,800 &   0.9   &  2.5       \\
\bottomrule
\end{tabular}
\end{table*}

\subsection{Photometry}
\subsubsection{K2 observations}\label{subec:k2_obs}
SDSS1238 was observed by the \textit{Kepler} Space Telescope during Campaign~10 of the \textit{K2} mission. It is designated EPIC\,228877473 ($K_p = 17.8\,$mag) in its Ecliptic Plane Input Catalogue. The Campaign~10 observations, taken between 2016 July 6 and September 20, suffered from an initial pointing error as well as the failure of one of the CCD modules on the spacecraft. As a result, the first six days of data were lost and the remaining observations have a 14-day gap during which the photometer powered itself down in response to the failure of Module \#4.

SDSS1238 was observed in short cadence mode (58.9\,s exposures) between July 13 -- 20 (MJD 57582.1 -- 57589.3) and then uninterrupted between August 3 and September 20 (MJD 57603.3 - 57651.2). It is the only object in the $5\times6$ pixel image stamp, so we summed the flux in a 12-pixel area, centred on the target, and used the remaining 18 pixels to estimate and subtract the background. We then removed times of thruster firings from the light curve, by comparing the centroid positions of the target in adjacent exposures. Deviations of more than $4\sigma$ were assumed to be due to the regular small adjustments that have to be made to the spacecraft position to correct for drift. SDSS1238 fell on a CCD close to the centre of the field, so experienced only a small amount of drift. We therefore did not apply the self-flat fielding procedure \citep{vdbj14} to the light curve, as it made little difference. Finally, we removed discrepant light curve points flagged with non-zero quality flags. The flags indicate that these were mostly due to cosmic ray strikes. We converted the summed counts $f$ to an approximate \textit{Kepler} magnitude $K_p$, using $K_p = 25.3 - 2.5 \log_{10}(f)$ \citep[e.g.][]{lund15,huber16}.

\subsubsection{Superoutburst: ground--based photometry}
On 2017 July 31 K.~Stanek reported the outburst (VSNET--alert~21308) of SDSS1238, detected by the All-Sky Automated Survey for Supernovae (ASAS--SN, \citealt{Shappee+2014,Kochanek+2017}). SDSS1238 brightened from a quiescent magnitude $V \simeq 17.8\,\mathrm{mag}$ to $V \simeq 11.86\,\mathrm{mag}$. This was the first outburst ever detected for this system. Following the detection of the outburst, an intensive monitoring was carried out by the global amateur community. Unfortunately the superoutburst occurred at the very end of the observational season of SDSS1238, when the star was only visible for about 1.5 hours after sunset from the southern hemisphere. Time--resolved photometry was obtained using seven telescopes around the world for about two weeks after the event. In order to maximise the data acquired, observations were often started in twilight and continued up to very high airmass (up to $\simeq 3$). In these cases, blue stars were used to compute the differential photometry in order to account for differential extinction. A summary of the observations is reported in Table~\ref{table:obs_log_superoutburst}.

\begin{figure*}
\includegraphics[width=0.9\textwidth]{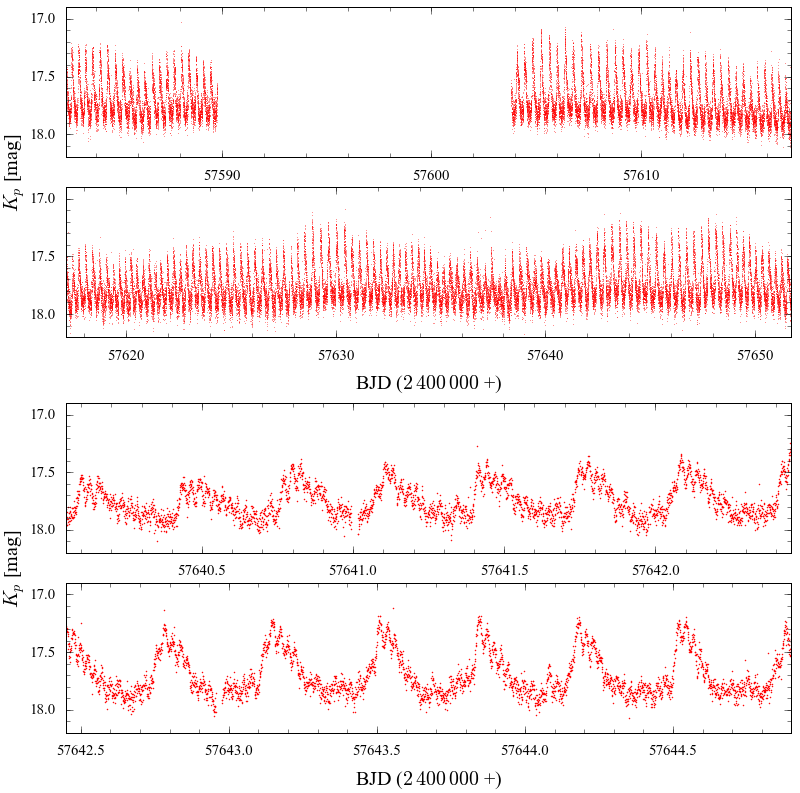}
\caption{\label{fig:k2_lc}The top two panels show the entire \textit{K2} light curve, the bottom two panels zoom in on a five-day segment. The 14\,d gap is related to the loss of CCD Module \#4. The \textit{K2} light curve is dominated by quasi-periodic brightenings recurring every $\simeq8.5$\,h with a typical amplitude of $\simeq0.5$\,mag, though a significant modulation of this amplitude is seen throughout the \textit{K2} Campaign. Superimposed on the $\simeq8.5$\,h brightenings is a photometric modulation with a coherent period of 40.26\,min, i.e. half the orbital period. The amplitude of these double-humps increases during the brightenings.}
\end{figure*}

\begin{figure*}
 \includegraphics[width=\textwidth]{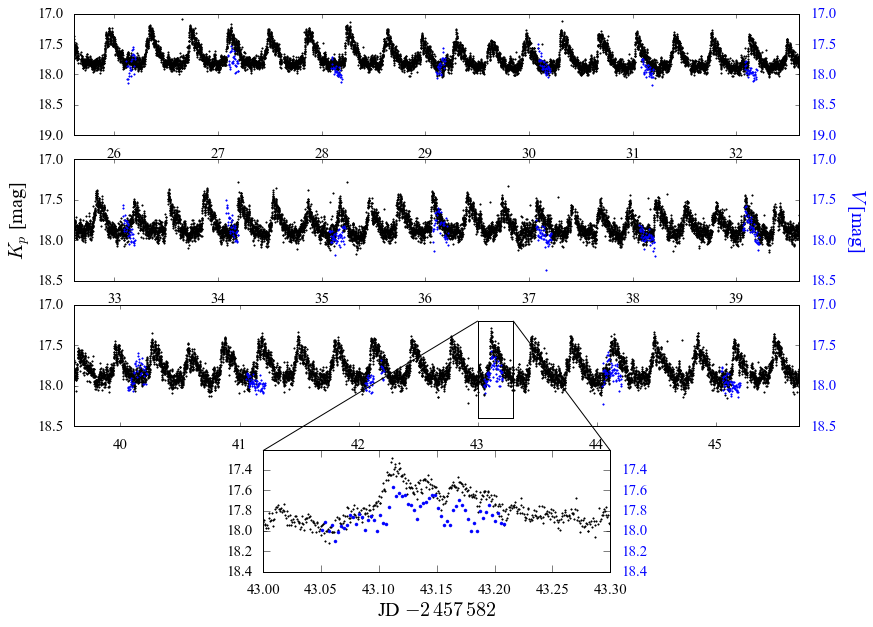}
 \caption{$K2$ light curve (black) from 2016 in comparison with $V$--band AAVSO observations (blue) acquired in a period of particular intense monitoring (20 February 2014 -- 13 March 2014). The AAVSO data are shifted in time by $\simeq 898.92\,$d to overlap with the \textit{K2} observations for an easy comparison. The quasi--period brightenings and the double--humps are clearly visible in both the datasets, showing the stability of both phenomena on timescales of years.}\label{fig:k2_aavso} 
\end{figure*}

\subsubsection{ULTRACAM observations}
High-time resolution observations were carried out on 2018 April 14 and 15 using ULTRACAM \citep{Dhillon+2007}, eight months after the superouburst (program ID 0101.D-0832). ULTRACAM is a high-speed CCD camera located at the Nasmyth focus of NTT in La Silla (Chile) and its triple beam setup allows to simultaneously acquire photometry in three different passbands. The observations of SDSS1238 were obtained using the Super-$u$, Super-$g$ and Super-$r$ filters (i.e. a set of custom filters covering the same wavelength range as the standard SDSS $ugr$ filters but with a higher throughput), with an exposure time of 10~s in each band. The Super-$u$ frames were co-added, i.e. combined before readout, in sets of three in order to achieve a higher signal--to--noise ratio (SNR).  A summary of the observations is reported in Table~\ref{table:obs_log_superoutburst}.

The data were reduced using the ULTRACAM pipeline. To account for variations in the observing conditions and to correct for the atmospheric extinction, the data were calibrated against a non-variable comparison star in the field view.  Finally, differential photometry was performed using the tabulated $ugr$ magnitudes of several comparison stars. Note that our calibration is subject to small systematic offsets because of the wider passbands of the Super SDSS filters compared to the normal $ugr$ filters. Nonetheless, this does not affect the timing analysis reported in Section~\ref{subsec:ultracam_analysis}.

\subsection{Spectroscopy}
\subsubsection{\hst/COS observations}\label{subec:hst_obs}
SDSS1238 was observed as a part of a large \hst programme in Cycle 20 (ID 12870) on 2013 March 1, using the Cosmic Origin Spectrograph (COS),

The data were collected over three consecutive spacecraft orbits for a total exposure time of 7183\,s through the Primary Science Aperture. The far--ultraviolet (FUV) channel, covering the wavelength range $1105\,$\AA\,$< \lambda < 2253\,$\AA, and the G140L grating, at the central wavelength of $1105\,$\AA\ and with a nominal resolving power of $R \simeq 3000$, were used. Since the detector sensitivity quickly decreases at the reddest wavelength, the useful wavelength range is reduced to $1150\,$\AA\,$ < \lambda < 1730\,$\AA.  The data were collected in \textsc{time--tag} mode, i.e. recording the time of arrival and the position on the detector of each detected photon, which also allows us the construction of ultraviolet light curves. 

\citet{survey_paper} report on the results of this large programme and showed that the \hst observations were carried out during the quiescent phase of SDSS1238. 

\subsubsection{VLT/X--shooter observations}\label{subec:xshoo_obs}
Phase--resolved spectroscopy of SDSS1238 was carried out on 2015 May 12 with X--shooter \citep{xshooter} located at the Cassegrain focus of UT2 of the VLT in Cerro Paranal (Chile).

X--shooter is an {\'e}chelle medium resolution spectrograph ($R \simeq 5000-7000$) consisting of three arms: blue (UVB, $\lambda \simeq 3000-5595\,$\AA), visual (VIS, $\lambda \simeq 5595-10\,240\,$\AA) and near--infrared (NIR, $\lambda \simeq 10\,240-24\,800\,$\AA). 

The X--shooter observations covered a whole orbital period of SDSS1238, with exposure times selected to optimise the signal--to--noise ratio (SNR) and, at the same time, to minimise the orbital smearing. We used the 1.0\arcsec, 0.9\arcsec\, and 0.9\arcsec\, slits for the UVB, VIS and NIR arm, respectively, in order to match the seeing, which was stable at $\simeq 1.0$\arcsec\, during the observations. The sky was not clear and some thick clouds passed over, consequently we had to discard two spectra for each arm owing to their extremely low SNR. Since the atmospheric dispersion correctors (ADCs) of X--shooter were broken at that time, the slit angle was reset to the parallactic angle position after one hour of exposure.

The data were reduced using the Reflex pipeline \citep{reflex}. To account for the well documented wavelength shift between the three arms\footnote{A report on the wavelength shift can be found at \url{https://www.eso.org/sci/facilities/paranal/instruments/xshooter/doc/XS\_wlc\_shift\_150615.pdf}}, we used theoretical templates of sky emission lines to calculate the shift of each spectrum with respect to the expected position. We then applied this shift when we corrected the data for the barycentric radial velocity.
 
Finally, a telluric correction was performed using \texttt{molecfit} \citep{molecfit1,molecfit2}. A log of the spectroscopic observations is presented in Table~\ref{table:obs_log}.

\begin{figure*}
 \includegraphics[width=0.85\textwidth]{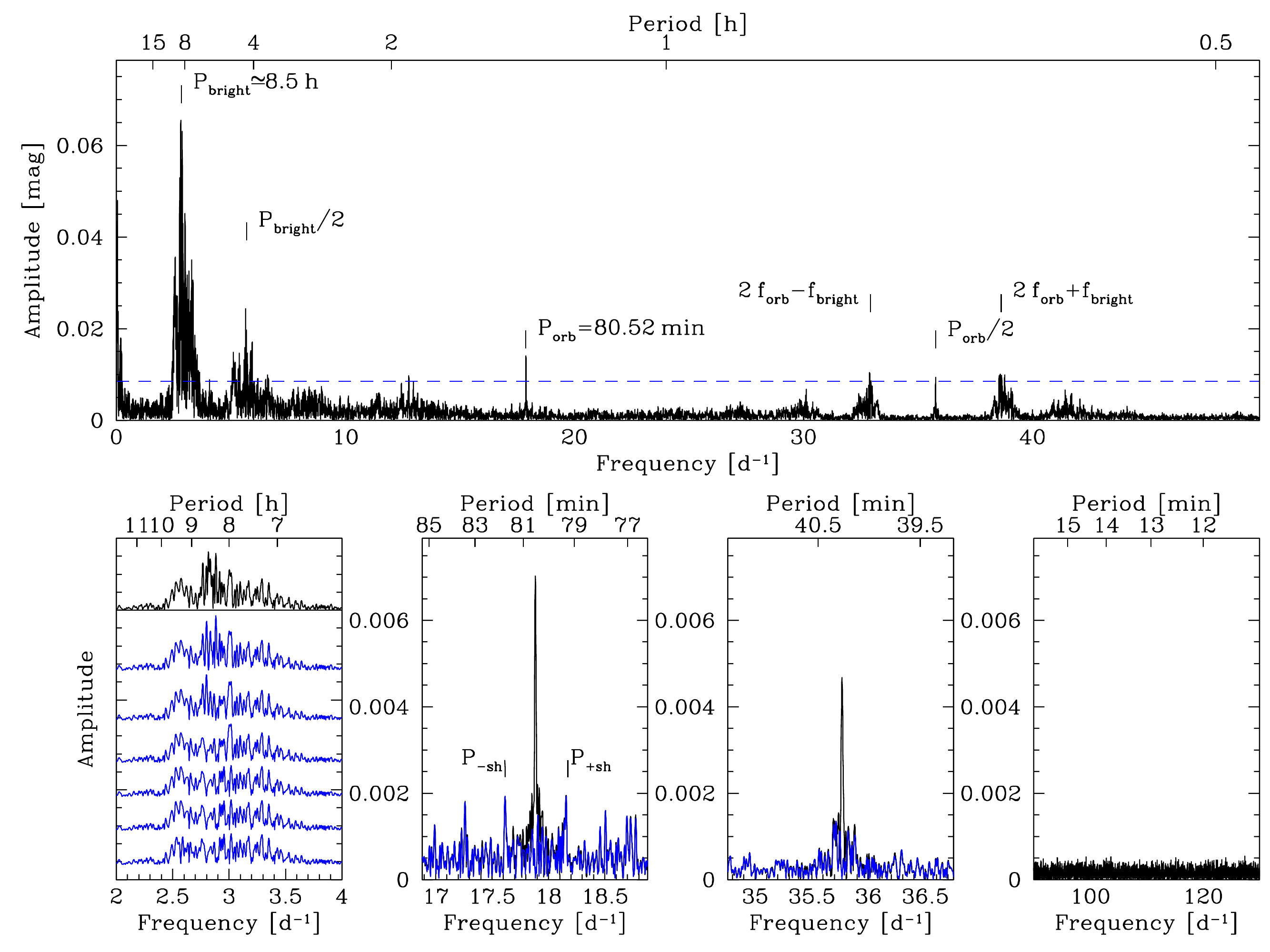}
 \caption{\label{fig:tsa} Top panel: amplitude spectrum of the full \textit{K2} light curve. The brightenings are non-coherent, resulting in a broad forest of signals centred on $\simeq8.5\,$h ($\simeq2.8\,\mathrm{d^{-1}}$). Subsequent pre-whitening of the \textit{K2} data with the strongest signal near $\simeq8.5$\,h (blue amplitude spectra on the bottom left panel) leaves significant structure, confirming the non-periodic nature of the brightenings. In contrast, the modulation at half the orbital period produces a sharp signal at $P_\mathrm{orb}$ and $1/2P_\mathrm{orb}$ in the amplitude spectrum, both signals can be cleanly removed (bottom middle left/middle right, blue curves). Sidebands are detected near the orbital period with a splitting that is consistent with positive and negative superhumps related to nodal and apsidal precession of a slightly eccentric disc. The \textit{K2} data do not show any evidence for the white dwarf spin period, $P_\mathrm{spin}=13.038\pm0.002$\,min (see Section~\ref{subsec:ultracam_analysis}).}
\end{figure*}

\begin{figure*}
 \includegraphics[width=\textwidth]{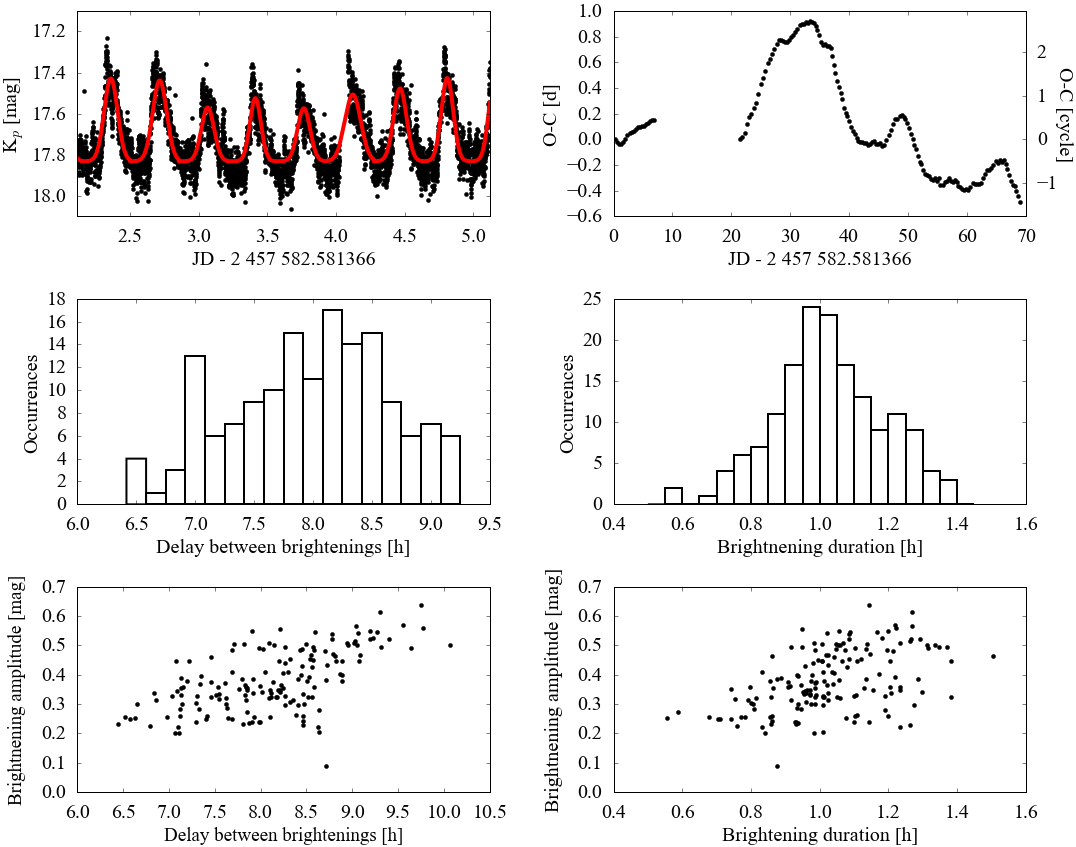}
 \caption{\textbf{Top.} \emph{Left:} Close--up of a short excerpt the \textit{K2} light curve (black) along with the best--fit model to the 8.5\,h brightenings (red). The full \textit{K2} light curve was fitted as the sum of a constant and 162 Gaussians. \emph{Right:} $O-C$ diagram for the brightenings observed in the \textit{K2} light curve. \textbf{Middle.} Histogram of the time difference between two consecutive brightenings (left) and the brightening durations (right). The mean and the standard deviations of these distributions are the average duration and recurrence time of the brightenings reported in the text. \textbf{Bottom.} Correlation between the amplitude and the delay between two consecutive brightenings (left) and between the amplitude and the brightening duration (right), which have Pearson coefficients of $\rho \simeq 0.6$ and $\rho \simeq 0.5$, respectively.}\label{fig:brightnening_properties} 
\end{figure*}

\begin{figure}
 \includegraphics[width=0.48\textwidth]{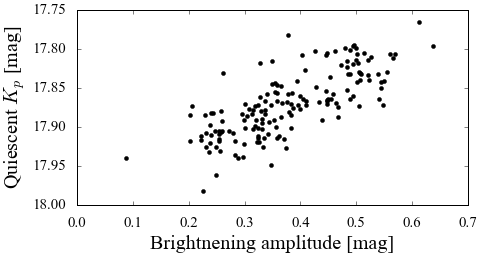}
 \caption{Correlation between the brightening amplitudes and the quiescent magnitude, calculated as the average magnitude between two consecutive brightenings.}\label{fig:amplitude_quiescence} 
\end{figure}

\section{Results}\label{sec:res}
\begin{figure*}
 \includegraphics[width=\textwidth]{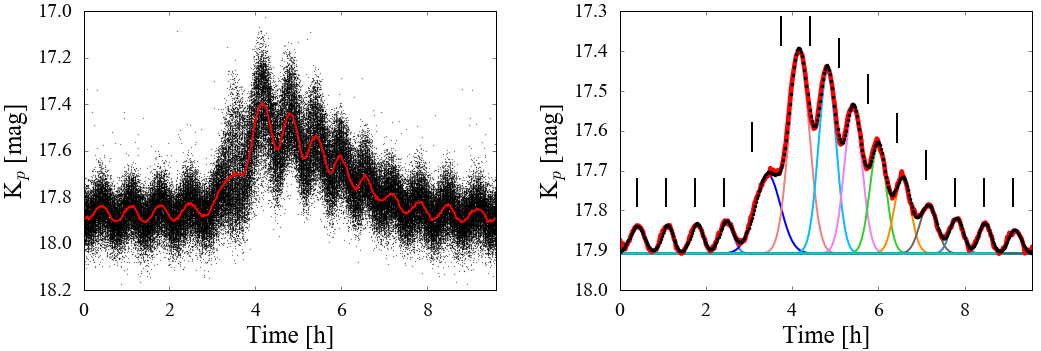}
 \caption{\label{fig:brightenings_phased} The 162 brightenings superimposed with respect to the 40.26\,min ephemeris of the double-humps (left, black dots). The double-humped structure is clearly coherent throughout the \textit{K2} campaign, and shows an increase in amplitude during the brightenings. The red line represents the data binned over an interval of five seconds. The binned light curve is shown again on the right (red line, note the different scale of the $y$--axis) along with the best fit model (black, dotted), which is composed of a series of 14 Gaussians of different widths and amplitudes. The black ticks repeat regularly every $40.26024$\,min to show that the occurrence of the humps is delayed (i.e. they undergo a phase shift) during the brightening.}\label{fig:kepler_shifted}
\end{figure*}

\begin{figure}
 \includegraphics[width=\columnwidth]{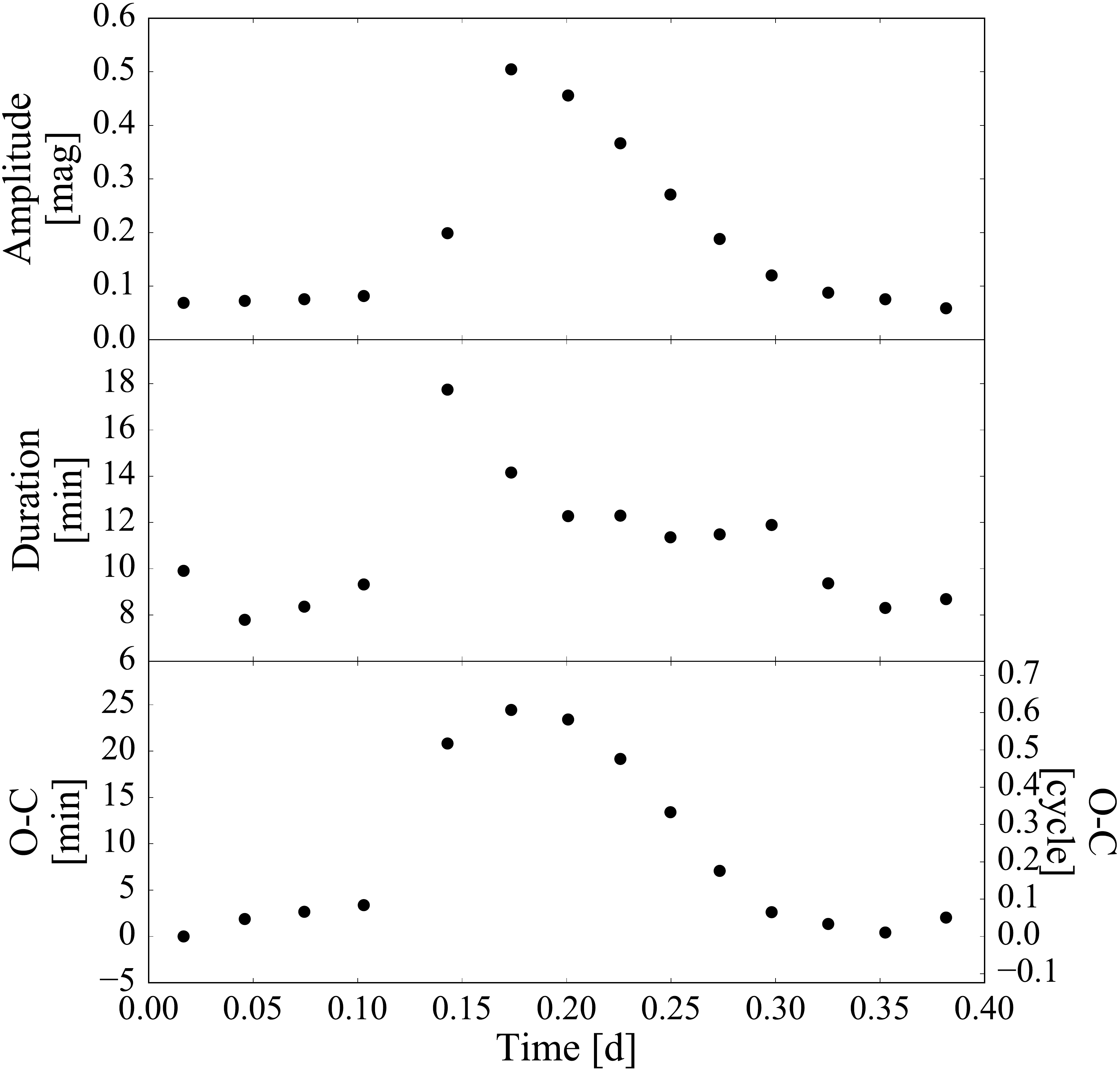}
 \caption{$O-C$ diagram (bottom panel) for the humps observed in the shifted \textit{K2} light curve as fitted by a series of 14 Gaussians (see Figure~\ref{fig:brightenings_phased}). During the brightening, the humps are phase shifted by about a half cycle. The hump durations and their amplitudes correlate with the occurrence of the brightenings (middle and top panel, respectively). During the brightenings, the humps last longer ($\delta t_\mathrm{hump} = 12.3 \pm 0.1\,$min) than in quiescence ($\delta t_\mathrm{hump} = 8.7 \pm 0.7\,$min) and have larger amplitudes ($\langle K_p \rangle = 0.46 \pm 0.06\,$mag during the brightening and $\langle K_p \rangle = 0.075 \pm 0.009\,$mag in quiescence).}\label{fig:oc_ampl_short} 
\end{figure}

\subsection{\textit{K2} quiescent light curve analysis}\label{subsec:k2_analysis}
The \textit{K2} light curve (Figure\,\ref{fig:k2_lc}) shows the two main characteristics previously identified in ground-based photometry (\citealt{Zharikov_2006,Aviles2010}, Figure~\ref{fig:k2_aavso}), i.e. quasi--periodic brightening events recurring every $\simeq8.5$\,h (Figure\,\ref{fig:k2_lc}, top panels), with a typical amplitude of $\simeq0.5$\,mag, and a humped photometric variability modulated at 40.26\,min, i.e. half the orbital period (Figure\,\ref{fig:k2_lc}, bottom, Figure~\ref{fig:k2_aavso}). The amplitude of the 8.5\,h brightenings varies on time scales of days throughout the $K2$ campaign, as can be seen in the top panels of Figure\,\ref{fig:k2_lc}. It is also evident that there are long-term variations in the overall brightness of the system, that modulate both the inter-brightnening magnitude and the amplitude of the brightenings on time scales of tens of days. The double--humps are persistently detected and they increase in amplitude during the brightenings.

We computed discrete Fourier transforms of the \textit{K2} data using the \textsc{MIDAS/TSA} package. As expected, the power spectrum is dominated by a forest of signals near $P_\mathrm{bright}\simeq8.5$\,h ($f_\mathrm{bright}\simeq2.8\,\mathrm{d}^{-1}$, Figure\,\ref{fig:tsa}, top panel), corresponding to the brightening events, and additional power is seen at the harmonic of that signal. Iteratively pre-whitening the \textit{K2} light curve with the strongest signal and re-computing a Fourier transform shows that multiple strong signals remain in the range $\simeq7-9$\,h ($\simeq2.5-3.5\,\mathrm{d}^{-1}$), underlining the incoherent nature of $P_\mathrm{bright}$ (Figure\,\ref{fig:tsa}, bottom left panel). In addition, the power spectrum contains two sharp signals at $\porb=80.52$\,min ($f_\mathrm{orb}\simeq17.9\,\mathrm{d}^{-1}$) and $1/2\,\porb=40.26$\,min. Both signals are cleanly removed by pre-whitening the \textit{K2} data (Figure\,\ref{fig:tsa}, bottom middle and right panel), as expected for the stable signal of the orbital period. Noticeable are the two sidebands seen at $2f_\mathrm{orb}\pm f_\mathrm{bright}$, which are the result of the strong amplitude modulation of the double-hump signal during the brightenings (see Section~\ref{subsec:humps}). There are also sidebands detected at the orbital frequency, with a splitting that agrees closely with the superhump period measured from the superoutburst photometry (Section\,\ref{subsec:superoutburst}), indicating that the disc maintains some slight eccentricity also in quiescence, resulting in nodal and apsidal precession. Finally, we do not detect any signature of the white dwarf spin $P_\mathrm{spin}=13.038 \pm 0.002$\,min (Section~\ref {subsec:system_par}). 

\subsubsection{Brightening analysis}\label{sec:brightening}
The strongest signal in the periodogram of the \textit{K2} light curve corresponds to a period of $\simeq8.5\,\mathrm{h}$ and arises from the 162 brightenings occurring during the \textit{Kepler} observations. This signal is quasi-periodic, but not strongly coherent (left lower panel of Figure~\ref{fig:tsa}). In order to study the recurrence time and the amplitude of these phenomena in more detail, we used a $\chi^{2}$ minimisation routine to fit the data with a sum of a constant value, to reproduce the quiescent flux, and 162 Gaussians to reproduce the shape of each individual brightening. 
The free parameters of this model are the quiescent level and the central time, the amplitude and the width of each Gaussian. A portion of the \textit{K2} light curve along with the best fit model is shown in the top left panel of Figure~\ref{fig:brightnening_properties}. 
Although the shape of the brightening is not strictly Gaussian, using a Gaussian model has the advantage of requiring a relatively small number of free parameters. Moreover, using a more complex model would introduce additional complexity in the definition of their central time. Equally, the  40.26\,min humps superimposed on the brightenings imply that simply defining the central time as the moment of maximum brightness will result in inconsistent results between different brightenings (Figure\,\ref{fig:k2_lc}).

From the width of each Gaussian we estimated the average brightening duration, $\delta t _\mathrm{bright}= 1.0 \pm 0.1\,\mathrm{h}$ (middle right panel of Figure~\ref{fig:brightnening_properties}). By measuring the delay time between two consecutive brightenings (middle left panel of Figure~\ref{fig:brightnening_properties}), we determined the average recurrence time of the brightenings, $t_\mathrm{rec}^\mathrm{bright} = 8.2 \pm 1.3\,\mathrm{h}$. Assuming this value, we calculated the $O-C$ diagram for the long period modulation (top right panel of Figure~\ref{fig:brightnening_properties}).

The smooth variation of the slope observed in the $O-C$ diagram implies that the brightnenings do not occur randomly but seem to evolve between a small number of preferred states. Further correlations are seen between their amplitudes, recurrence times and durations: the longer the delay between a brightening and its preceding, the larger its amplitude and the longer it lasts (bottom panels of Figure~\ref{fig:brightnening_properties}).
An additional correlation is found between the brightening amplitudes and the quiescente magnitude of the system (Figure~\ref{fig:amplitude_quiescence}). These correlations highlight the presence of a systematic mechanism acting behind the brightenings, whose origin will be discussed in detail in Section~\ref{sec:discussion}.

\begin{figure*}
 \includegraphics[angle=180,width=0.8\textwidth]{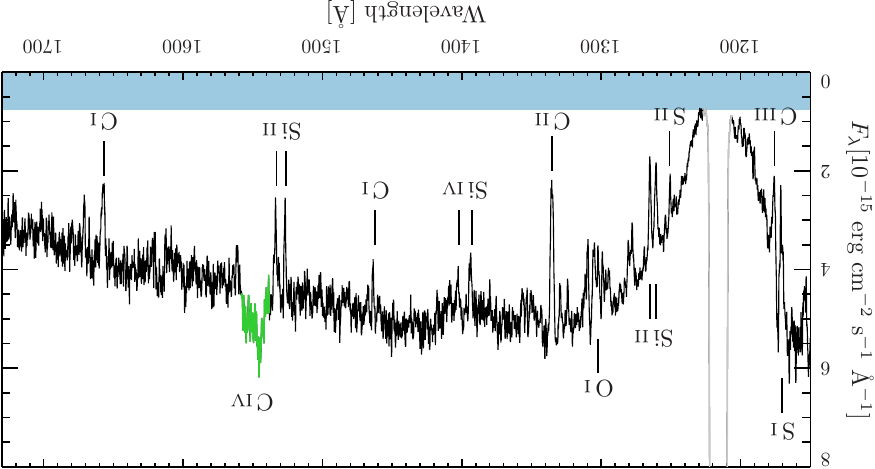}
 \caption{\textit{HST}/COS average spectrum of SDSS1238. The unmistakable white dwarf signature is the broad Ly$\alpha$ absorption, the quasi--molecular absorption band of $H_{2}^{+}$ at $\lambda \simeq 1400\,$\AA\, and several sharp photospheric metal absorption lines. Plotted in green is the \ion{C}{iv} ($1550\,$\AA) emission line originating in the accretion disc. The blue band illustrates the presence of a second continuum component which contributes $\simeq 15$ per cent of the observed flux and which we assume to be flat in $F_\lambda$. The geocoronal Ly$\alpha$ emission line ($1216\,$\AA) is plotted in grey.}\label{fig:hst_ave} 
\end{figure*}

\subsubsection{Double--hump analysis}\label{subsec:humps}
The coherent $40.26024 \pm 0.00060$\,min signal detected in the Fourier transform of the $\textit{K2}$ data, where the uncertainty was determined from a sine fit to the \textit{K2} data, which we dubbed ``double--humps'', is consistent with half the orbital period as measured from time-resolved spectroscopy \citep{Zharikov_2006}. In order to investigate possible correlations between the 8.5\,h brightenings and the double--humps, we divided the \textit{K2} light curve into 9.6\,h long chunks, each of them centred on the peak of one of the 162 brightening events (as measured from the Gaussian fits described in Sect.\,\ref{sec:brightening}), and shifted the start time of each chunk so that their peaks align. These shifts in time were carried out in steps of 40.26025\,min, so that the double-humps also align in phase. The resulting average light curve reveals that the brightenings repeat remarkably well in time, with a fast rise and a slower decline, and that the double--humps grow in amplitude during the brightenings (Figure\,\ref{fig:kepler_shifted}, right panel).  

As for the analysis of the brightenings outlined in the previous section, we fitted the double--humps in the average brightening light curve with a model which is the sum of 14 Gaussians (one for each hump detected) and a constant, to reproduce the quiescence level. The free parameters of the model are hence the quiescent magnitude, the amplitude, the width and the central time of the 14 Gaussians describing the humps. The best--fit model to the average brightening light curve is shown in the right panel of Figure~\ref{fig:kepler_shifted}. This model reproduces the average brightening remarkably well, without the need for any additional contribution beyond the Gaussians representing the individual 40.26\,min humps. In other words, this model provides the alternative interpretation that the brightenings and the double--humps are actually the same phenomenon, with the first ones being the superposition of the second ones. In this scenario, the double--humps significantly grow in amplitude every $\simeq 8.5\,$h, giving rise to the 162 brightenings detected in the \textit{K2} light curve. We discuss this possibility in more detail in Section~\ref{brightening_discussion}.

By measuring the time difference between two consecutive humps, we calculated the $O-C$ diagram for the double--humps, which is shown in the bottom panel of Figure~\ref{fig:oc_ampl_short}. During the brightening, the humps undergo a phase shift, i.e. their occurrence is delayed when the system brightens, as can be seen by the sharp increase and subsequent smooth decrease of the $O-C$ values. The widths and the amplitudes of the best--fit Gaussians strongly correlate with the occurrence of the brightenings. During the brightenings, the humps last, on average, longer ($12 \pm 1\,$min) than during quiescence ($8 \pm 1\,$min) and their amplitude varies from a quiescent value of $\langle K_p \rangle \simeq 0.05 \,$mag to $\langle K_p \rangle \simeq 0.5 \,$mag during the brightening (top panel of Figure~\ref{fig:oc_ampl_short}).

\subsection{Ultraviolet spectroscopy and light curve}\label{subsec:hst_analysis} 
Figure~\ref{fig:hst_ave} shows the average ultraviolet spectrum of SDSS1238 from the \hst/COS observations, obtained by summing the data from the three orbits. The emission is dominated by the white dwarf, recognisable in the broad Ly$\alpha$ absorption profile at $1216\,$\AA. Some contribution from the disc can be identified in the \ion{C}{iv} emission at $1550\,$\AA. Moreover, the non--zero flux in the core of the Ly$\alpha$ unveils the presence of an additional continuum component, which contributes $\simeq 15$ per cent of the observed flux\footnote{The origin of this additional emission source is unclear and it has been suggested that it could arise from (i) the disc, or (ii) the region of intersection between the ballistic stream and the disc or (iii) the interface region between the disc and the white dwarf surface, see \citet{survey_paper} for a more detailed discussion.}. From a fit to this spectrum, \citet{survey_paper} determined an effective temperature of the white dwarf of $\twd = 18\,358 \pm 922\,\mathrm{K}$ for $\log g = 8.35$ and $Z = 0.5\,Z_\odot$. 

The \textsc{time--tag} data allow us to reconstruct a 2D image of the detector, where the dispersion direction runs along one axis and the spatial direction along the other. In order to extract the light curve, we identified three rectangular regions on this image: a central one enclosing the object spectrum and two regions to measure the background, one above and one below the spectrum. We masked the geocoronal emission lines from Ly$\alpha$ ($1206\,$\AA\,$ < \lambda < 1225\,$\AA) and \ion{O}{i} ($1297\,$\AA\,$ < \lambda < 1311\,$\AA) and the \ion{C}{iv} emission from the accretion disc ($1543\,$\AA\,$< \lambda < 1556\,$\AA). Finally, we binned the data into five second bins.

\begin{figure*}
 \includegraphics[width=0.8\textwidth]{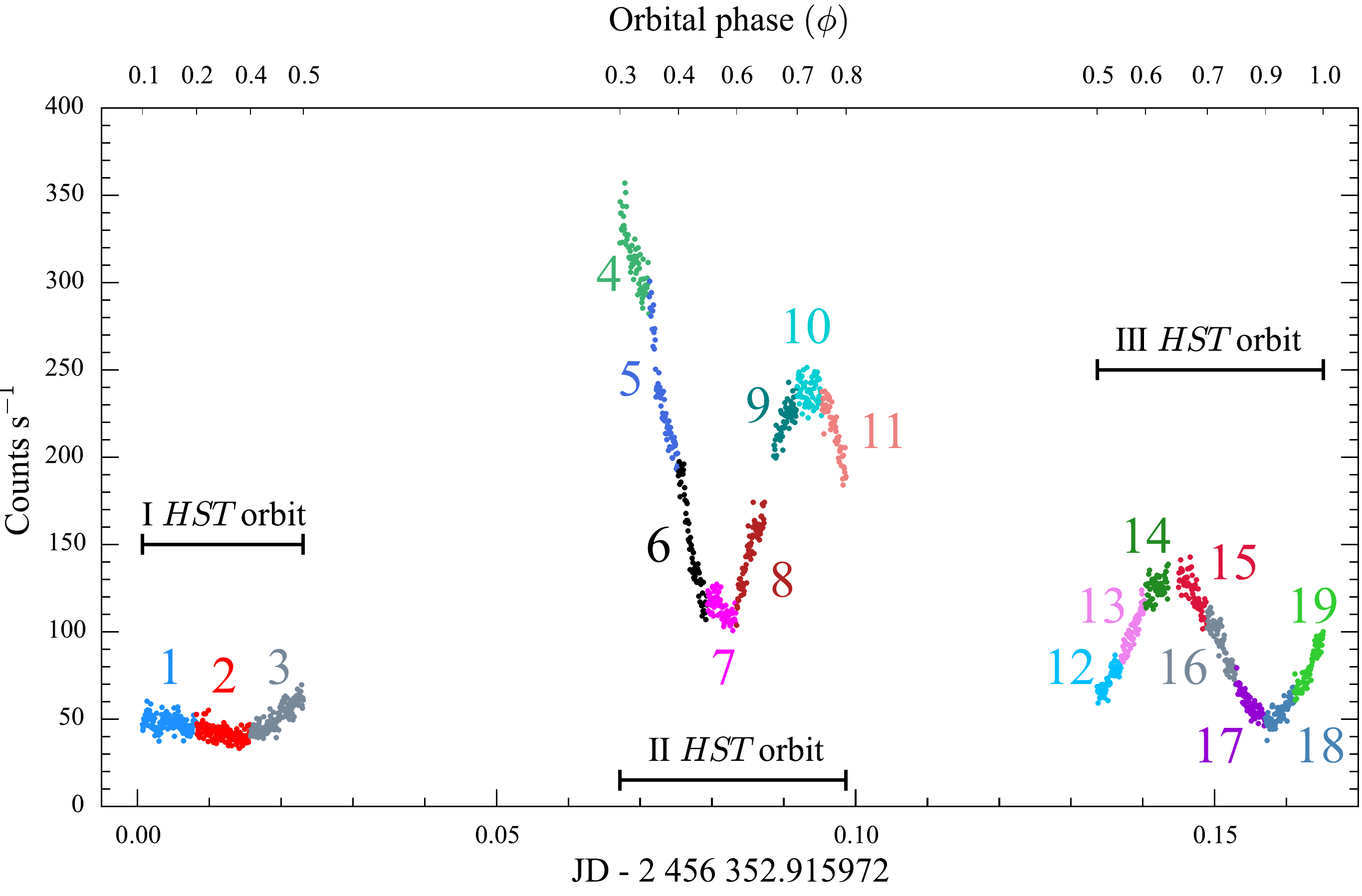}
 \caption{\textit{HST}/COS light curve of SDSS1238, obtained over three consecutive spacecraft orbits, the gaps result from the target being occulted by the Earth. During the observations, the system showed an increase in brightness of $\simeq 2\,$mag superimposed by a smaller amplitude ($\simeq 1\,$mag) sinusoidal variation on time scales of half the orbital period. The colours highlight the different exposures from which the 19 sub--spectra have been extracted.}\label{fig:hst_lightcurve} 
\end{figure*}

The COS light curve (Figure~\ref{fig:hst_lightcurve}) shows considerable variability of the ultraviolet flux, a large increase in the overall brightness of $\simeq 2\,$mag super--imposed by a smaller amplitude ($\simeq 1\,$mag) sinusoidal variation on time scales close to half of the orbital period. Therefore, the spectrum in Figure~\ref{fig:hst_ave}, obtained by averaging over the three spacecraft orbits, is only representative of a time--averaged temperature of the white dwarf, and a more detailed analysis is required to investigate the temporal evolution of the system. We used the splitcorr routine from the \textsc{calcos} COS pipeline to split the observations into 19 shorter exposures, from which we extracted the corresponding spectra using the \textsc{calcos} x1dcorr routine.

\begin{figure}
 \includegraphics[width=0.48\textwidth]{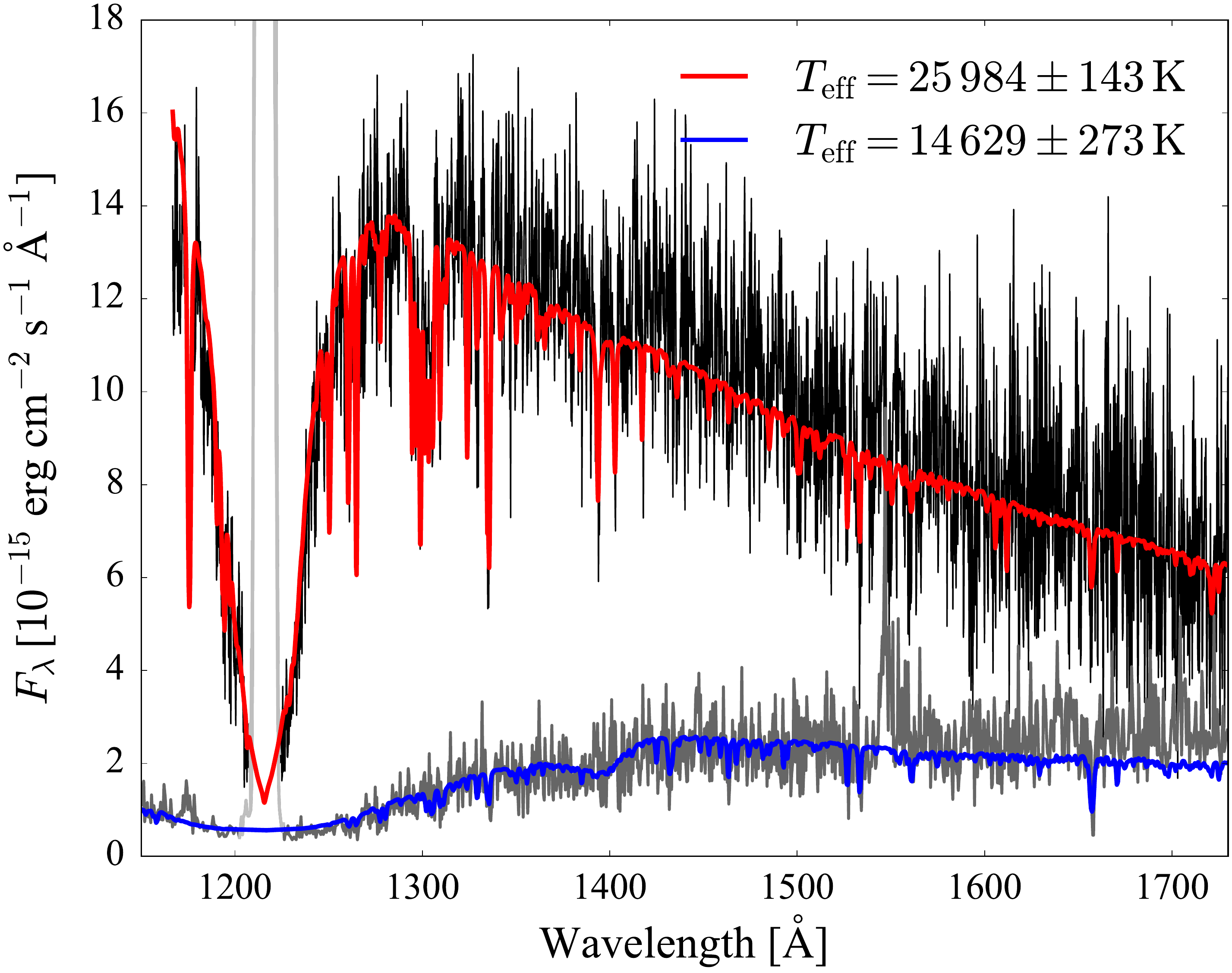}
 \caption{\textit{HST}/COS spectra of SDSS1238 extracted at minimum (\#18, dark grey) and at maximum (\#4, black) brightness level (see Figure\,\ref{fig:hst_lightcurve}) along with the best fit models (blue, $T_\mathrm{eff}=14\,629$\,K and red, $T_\mathrm{eff}=25\,984$\,K, respectively). The geocoronal Ly$\alpha$ ($1216\,$\AA) emission lines are plotted in light grey.}\label{fig:hst_spectra} 
\end{figure}

Figure~\ref{fig:hst_spectra} shows the \hst spectra corresponding to the minimum (\#18, blue) and the maximum (\#4, red) ultraviolet flux of SDSS1238. The spectral shape varies from that of a cool white dwarf (\#18, with a very broad Ly$\alpha$ absorption extending to $\simeq 1600\,$\AA) to that of a hot white dwarf (\#4, with a narrow Ly$\alpha$ profile and a steep blue continuum), revealing the heating and subsequent cooling of the white dwarf in SDSS1238.

We carried out spectral fits to the ultraviolet spectra to  study  the evolution of the white dwarf effective temperature throughout the \textit{HST}/COS observations. We generated a grid of white dwarf synthetic models using \textsc{tlusty} and \textsc{synspec} \citep{tlusty,tlusty1}, covering the effective temperature range $T_\mathrm{eff} = 10\,000 -  40\,000\,$K in steps of $100\,$K, for $Z = 0.5\,\mathrm{Z}_\odot$ (as measured by \citealt{survey_paper} from the analysis of the average \hst spectrum). 
Assuming the distance to the system $d = 170^{+5}_{-4}\,$pc as derived from the \textit{Gaia} parallax ($\varpi = 5.9 \pm 0.2\,$mas, \citealt{Gaia2016,Gaia2018}), we fitted the \hst data and determined the white dwarf effective temperature for each of the 19 sub--spectra. To account for the presence of the additional continuum component, we included a constant flux (in $F_\lambda$) in the fit. Although different approximations can be used to model this second component (such as a power law or a blackbody, see \citealt{survey_paper} for a detailed discussion on this topic), we chose to use a constant flux since as it is the simplest model and it introduces only one additional free parameter in the fit. Hereafter, we refer to this fitting method as the two--component fit (one white dwarf plus a constant).

We performed the spectral fit using the Markov chain Monte Carlo (MCMC) implementation for Python, \textsc{emcee} \citep{emcee}. We assumed a flat prior for the white dwarf effective temperature in the range $9000\,\mathrm{K} < T_\mathrm{eff} < 40\,000\,\mathrm{K}$ and constrained the white dwarf scaling and the additional second component to be positive. The best--fit results are summarised in Table~\ref{table:temp_var}. 

\begin{table}
\caption{Effective temperature variation of the white dwarf in SDSS1238.}\label{table:temp_var}
\setlength{\tabcolsep}{0.1cm}
\begin{center}
\begin{tabular}{lcccccc}  
\toprule
            &          & \multicolumn{2}{c}{Two--component fit} & & \multicolumn{2}{c}{Three--component fit}\\\cmidrule{3-4}\cmidrule{6-7}
Spectrum    & Exposure & $\twd$  & $\pm$ & & $T_\mathrm{hot}$ & $\pm$  \\
            & time (s) &  (K)    &  (K)  & &   (K)            &  (K)  \\
\midrule
\# 1         &  648     & 15\,101   & 245    &&   16\,371   & 768 \\
\# 2         &  648     & 14\,892   & 226    &&   16\,606   & 1014 \\
\# 3         &  648     & 15\,879   & 219    &&   16\,578   & 357 \\
\# 4         &  350     & 25\,984   & 143    &&   25\,572   & 235 \\
\# 5         &  350     & 25\,489   & 277    &&   25\,072   & 262 \\
\# 6         & 350      & 23\,907   & 257    &&   23\,351   & 241 \\
\# 7         & 350      & 20\,862   & 330    &&   21\,256   & 312 \\
\# 8         & 350      & 20\,603   & 272    &&   20\,568   & 210 \\
\# 9         & 293      & 21\,590   & 268    &&   21\,358   & 202  \\
\# 10       & 293      & 21\,916   & 279.   &&   21\,627   & 215  \\
\# 11       & 293      & 21\,888   & 274    &&   21\,620   & 197 \\
\# 12       & 288      & 15\,415   & 296    &&  16\,965    & 972  \\
\# 13       & 288      & 16\,799   & 254    &&  17\,181    & 340 \\
\# 14       & 288      & 17\,618   & 253    &&  17\,895    & 289 \\
\# 15       & 350      & 18\,341   & 243    &&  18\,656    & 235 \\
\# 16       & 350      & 17\,957   & 241    &&  18\,468    & 340 \\
\# 17       & 350      & 15\,626   & 319    &&  17\,273    & 872 \\
\# 18       & 350      & 14\,629   & 273    &&  16\,579    & 1403 \\
\# 19       & 350      & 15\,975   & 233    &&  16\,502    & 411 \\
\bottomrule
\end{tabular}
\end{center}
\end{table}

\begin{figure*}
 \includegraphics[width=0.8\textwidth]{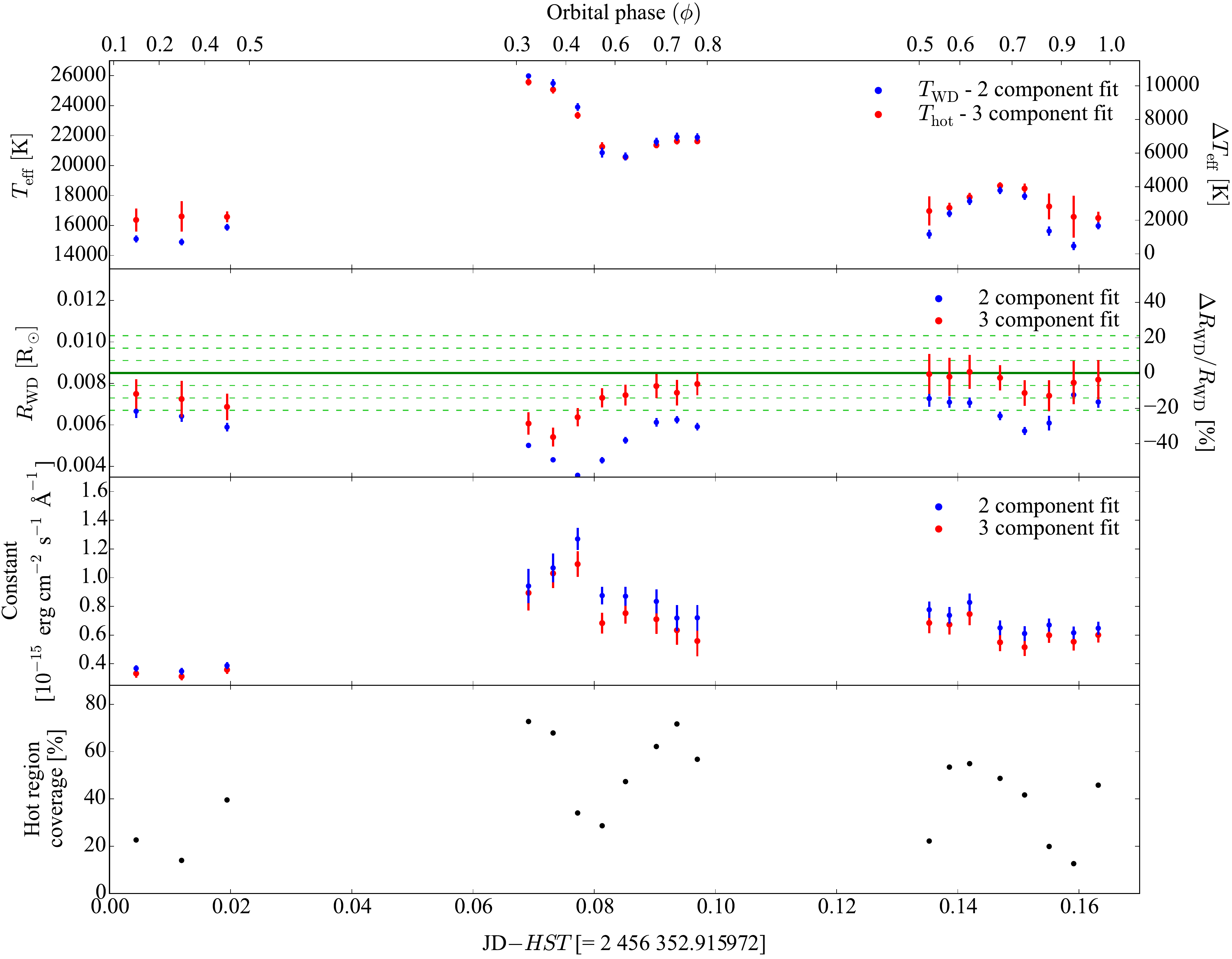}
 \caption{Best--fit results for the 19 \hst sub--spectra for the two--component fit (blue dots) and the three--component fit (red dots). In the second panel, the solid green line represents the white dwarf radius as determined from the analysis of the X--shooter data (see Section~\ref{subsec:system_par}). The dot--dashed green lines are the relative 1$\sigma$, 2$\sigma$ and 3$\sigma$ uncertainties.
The strong change of the (apparent) white dwarf radius in the two--component fit (blue dots) rules this model out as a physically meaningful description. The bottom panel shows the fraction of the visible white dwarf surface covered by the hotter component.}\label{fig:fit} 
\end{figure*}

The scaling factor between each sub--spectrum and the related best-fit model corresponds to an \textit{apparent} white dwarf radius, according to the equation:
\begin{equation}\label{eq:scaling}
S = \left( \frac{R_\mathrm{WD}}{d} \right) ^2
\end{equation}
which varies from $\simeq 0.004\,\rsun$ to $0.009\,\rsun$ as a function of the heating and cooling (Figure~\ref{fig:fit}, blue points, second panel).

This $\simeq 50$ per cent variation in the apparent radius implies that the heating occurs only over a fraction of the visible white dwarf surface. To account for the presence of a hotter region, we repeated the fit to the 19 \hst sub--spectra including two white dwarf models. In this procedure, we assumed an underlying white dwarf with $T_\mathrm{cool} = 14\,000\,\mathrm{K}$, slightly cooler than the minimum temperature measured from spectrum \#18 ($T_\mathrm{eff} = 14\,629\,\mathrm{K}$), and allowed the presence of an additional hotter white dwarf. We assumed a flat prior on its effective temperature, constraining its value in the range $T_\mathrm{cool} < T_\mathrm{hot} < 40\,000\,\mathrm{K}$. As before, we included an additional flat $F_\lambda$ continuum component. We constrained the flux scaling factors for the two white dwarf models and the $F_\lambda$ component to be positive. Following the same nomenclature as before, we refer to this prescription as three--component fit (two white dwarfs plus a constant).

We found that this three--component fit  successfully describes the observed variation of the ultraviolet flux, and the accompanying change of the white dwarf temperature, by heating and cooling of a region on the white dwarf surface. The fact that the ultraviolet flux is modulated at half the orbital period, just as the double--humps in the \textit{K2} light curve, actually requires the presence of \textit{two} hot regions on the white dwarf which rotate in and out of view during the orbital motion of the binary. The area covered by each region varies from $\simeq 5$ per cent, in quiescence, up to $\simeq 80$ per cent of the visible white dwarf surface. The flux contribution of the flat $F_\lambda$ component remains almost unchanged between the two and the three--component fits (Figure~\ref{fig:fit}, third panel).  

\begin{figure*}
 \includegraphics[width=\textwidth]{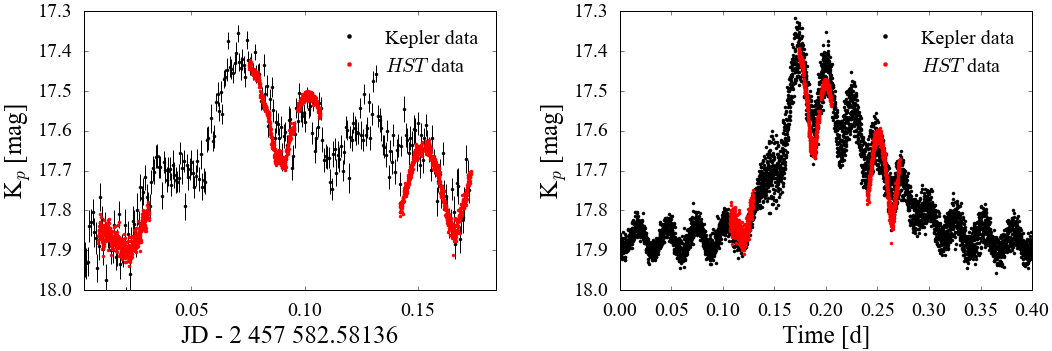}
 \caption{\hst light curve (red) in comparison with the \textit{K2} data (black) of a sample brightening (left) and with the average \textit{K2} light curve folded with respect to the 40.26\,min ephemeris of the double-humps (right). The \hst counts are converted into instrumental magnitudes and then scaled and shifted in time to overlap with the \textit{K2} observations. The close agreement between the shape of the two datasets demonstrates that the brightenings and humps detected in the optical light curve arise from the heating and the cooling of the white dwarf, and in addition illustrates the stability of both signals on time-scales of years. 
\label{fig:lc_kepler_hst}}
\end{figure*}

In the three-component fit, the apparent radius of the white dwarf (red points in Figure~\ref{fig:fit}, second panel) can be calculated using Equation~\ref{eq:scaling}, where $S$ is the sum of the two scaling factors of the cold and hot white dwarf models. In contrast to the two-component fit, the apparent white dwarf radius is in a better agreement with the value measured from the X--shooter data (see Section~\ref{subsec:system_par}).
Since the three-component fit does not imply a significant variation in the white dwarf size, we consider this a more physical and realistic model for the ultraviolet variability of SDSS1238. However, the presence of a residual variation in the white dwarf size also in three-component fit shows that this model is oversimplified, implying a discontinuity in the temperature profile at the edge of the hot region. A more accurate model should include a gradient in temperature over the whole visible white dwarf surface but such a more detailed model goes beyond the scope of this work.

\subsection{Energetics of the brightenings}\label{subsec:energetic}
\begin{figure*}
 \includegraphics[width=\textwidth]{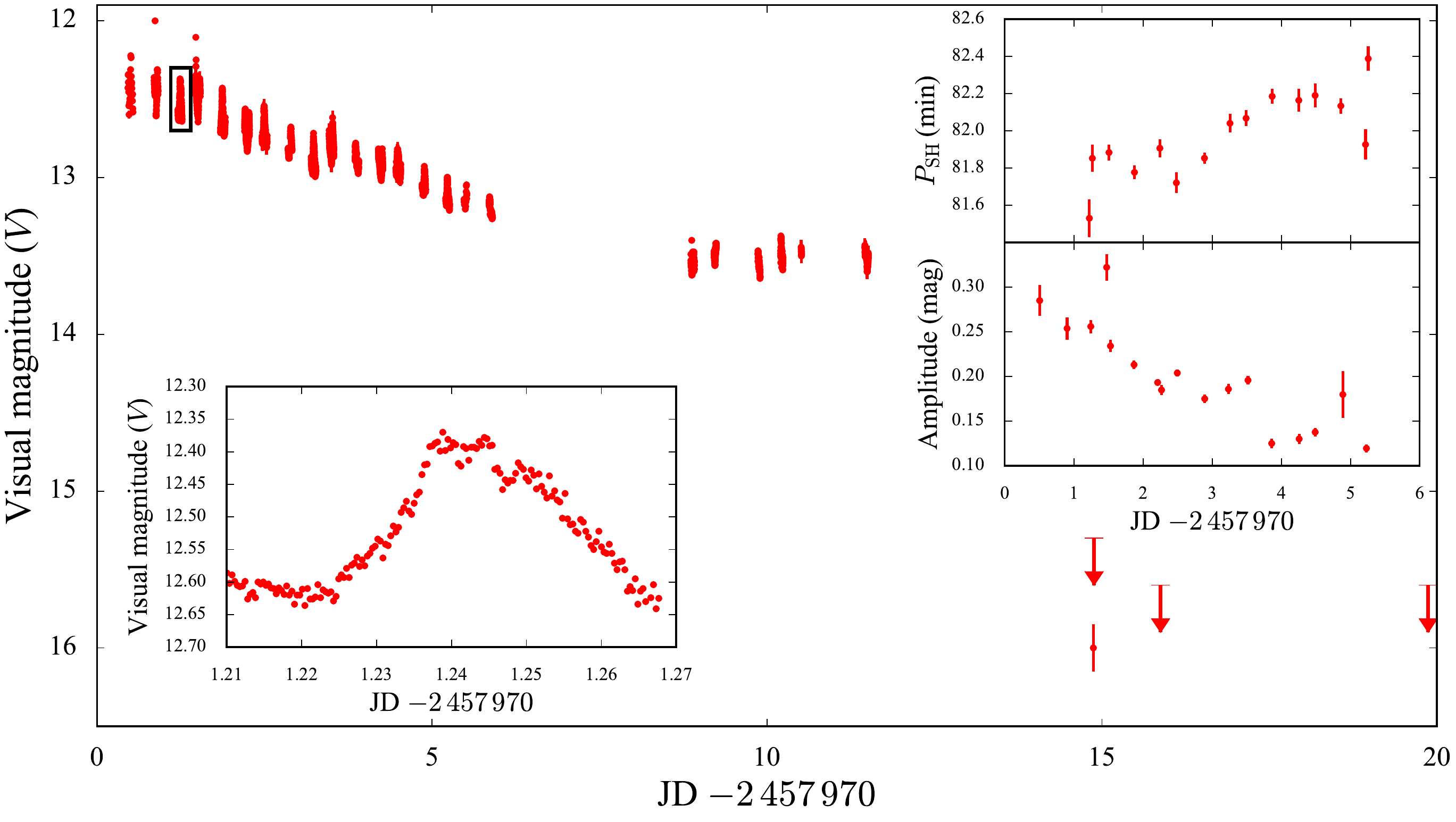}
 \caption{Superoutburst light curve of SDSS1238. The bottom left inset shows a close--up of the region highlighted with the black rectangle in the main plot; the top right insets show the $P_\mathrm{SH}$ (top) and the amplitude (bottom) variation of the superhumps detected before the observing gap.}\label{fig:superoutburst_light curve} 
\end{figure*}

The photometric variability of SDSS1238 observed in the ultraviolet closely resembles the shape of the optical light curve. In order to compare the \hst and \textit{K2} observations, we converted the \hst count rate light curve shown in Figure\,\ref{fig:hst_lightcurve} into magnitudes and then scaled it by a factor $0.2$ and shifted it in time in order to overlay them on the \textit{K2} data (Figure~\ref{fig:lc_kepler_hst}). The close agreement between the \hst and \textit{K2} dataset seen in Figure~\ref{fig:lc_kepler_hst} clearly demonstrates that the heating and the cooling of the white dwarf detected in the ultraviolet is also the origin of both the brightenings and the double--humps detected in the optical light curve of SDSS1238.

To establish the contribution of the (heated) white dwarf to the optical emission of SDSS1238, we convolved the best--fit models obtained from the three--component fit to the \hst data with the $K_p$ filter passband, obtaining the corresponding optical magnitudes of the white dwarf. While the ultraviolet emission is dominated by the white dwarf, its contribution in the optical varies from $\simeq 40$\,per cent in quiescence up to $\simeq 60$\,per cent during the brightening. The accretion disc makes up for the remaining half of the emission, in agreement with what is observed in the X--shooter SED, where the white dwarf contributes $\simeq 66\,$per cent and the disc contributes for $\simeq 34$\,per cent of the emission in the $K_p$ passband (Section~\ref{subsec:system_par}).

Finally, to estimate the energy excess associated with the brightenings, with respect to the quiescent level, we first calculated the average counts associated with each \hst sub-spectrum, scaling the value according to the percentage of white dwarf contribution with respect to the additional second component. We found an analytical expression for the relationship between the average counts and the effective temperature of each \hst sub-spectrum. Using the \hst data as a reference, we converted the average \textit{K2} light curve (see Section~\ref{subsec:humps}) in count/s. We then used the relation between the average \hst counts and effective temperature to calibrate the \textit{K2} data and, from the knowledge of the temperature, we computed the luminosity. 

The quiescent level, which we approximated with a constant, corresponds to a luminosity of $\simeq 1.1 \times 10^{31}\,\mathrm{erg/s}$ and arises from the compressional heating of the white dwarf by the accreted material, providing a direct measurements of the secular time-averaged accretion rate ($\langle \dot{M} \rangle$) onto the white dwarf \citep{Dean_and_Boris2009}. Using equation~1 from \cite{Dean_and_Boris2009}, we found that the quiescent luminosity corresponds to $\langle \dot{M} \rangle \simeq 4 \times 10^{-11}\,\mathrm{M}_\odot \mathrm{yr}^{-1}$, a typical value for a CV at the period minimum \citep{Dean_and_Boris2009,survey_paper}.

After subtracting the quiescent luminosity, the energy excess released during the average brightening corresponds to $\simeq 2 \times 10^{35}\,$erg. Assuming that the brightenings originate from accretion events, this energy excess would correspond to an accreted mass of $\simeq\,7\times\,10^{17}\,\mathrm{g}$  ($\simeq 4\times 10^{-16}\,\mathrm{M}_\odot$). The accretion rate onto the white dwarf during a typical brightening would hence result enhanced by $\dot{M} \simeq 7\times 10^{-13}\,\mathrm{M}_\odot \mathrm{yr}^{-1}$, with respect to the secular time-averaged quiescent accretion rate.

\begin{figure*}
  \includegraphics[width=\textwidth]{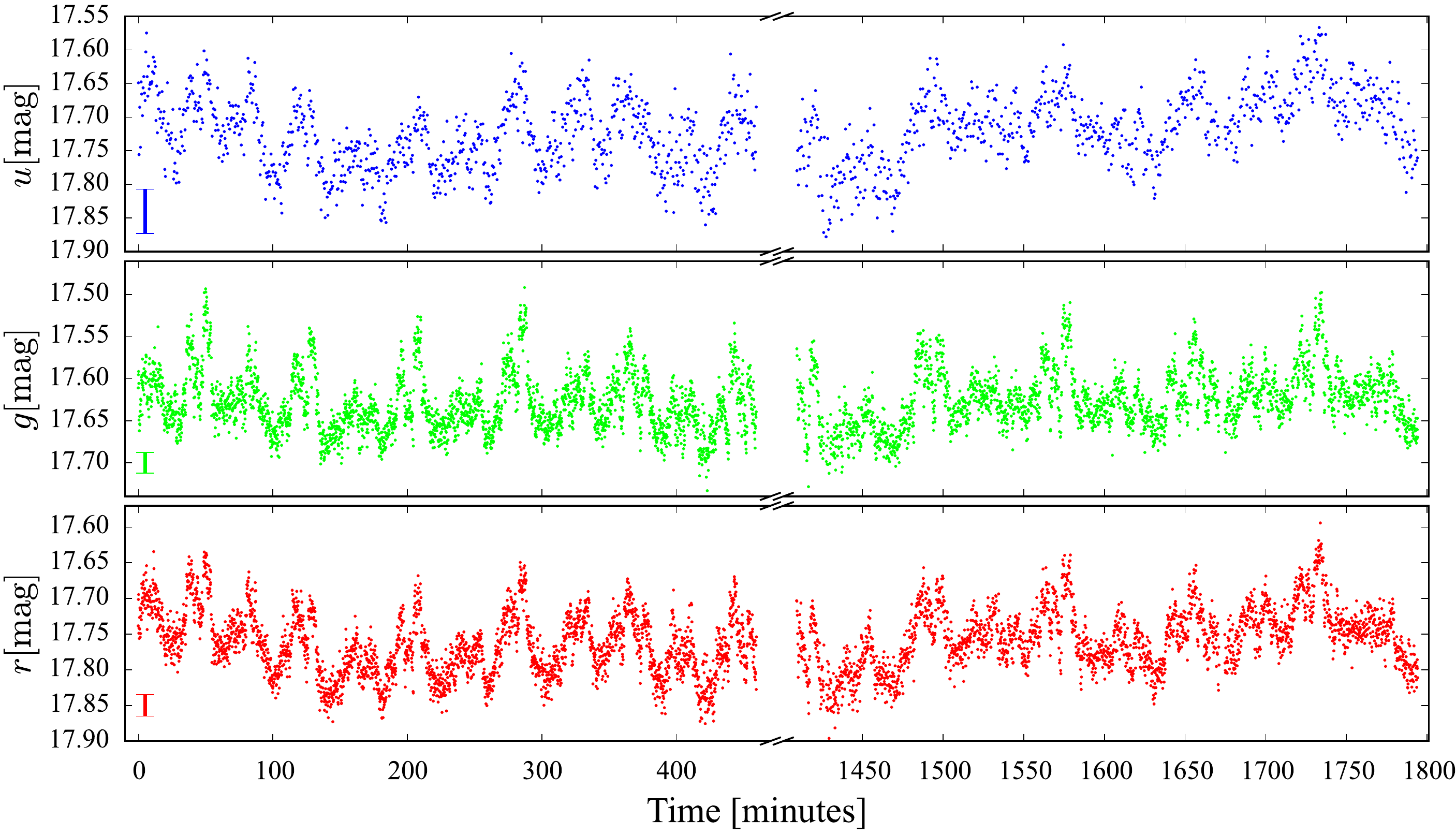}
 \caption{ULTRACAM light curves of SDSS1238 obtained eight months after the superoutburst, in the Super-$u$ (blue), Super-$g$ (green) and Super-$r$ (red) filters. The typical uncertainties on the measurements are shown by the errorbar on the bottom left of each subplot.
As revealed by the period analysis of the data (Figures~\ref{fig:power_ugr_k2_wide} and \ref{fig:fold_ugr}), the complex shape of the light curve is determined by the combination of two signals, one arising from the double humps previously detected in the $K2$ light curve, and an additional signal at $\simeq 13\,$min, which we interpreted as the spin of the white dwarf.}\label{fig:ultracam_lc}  
\end{figure*}

\begin{figure*}
    \centering
    \includegraphics[width=\textwidth]{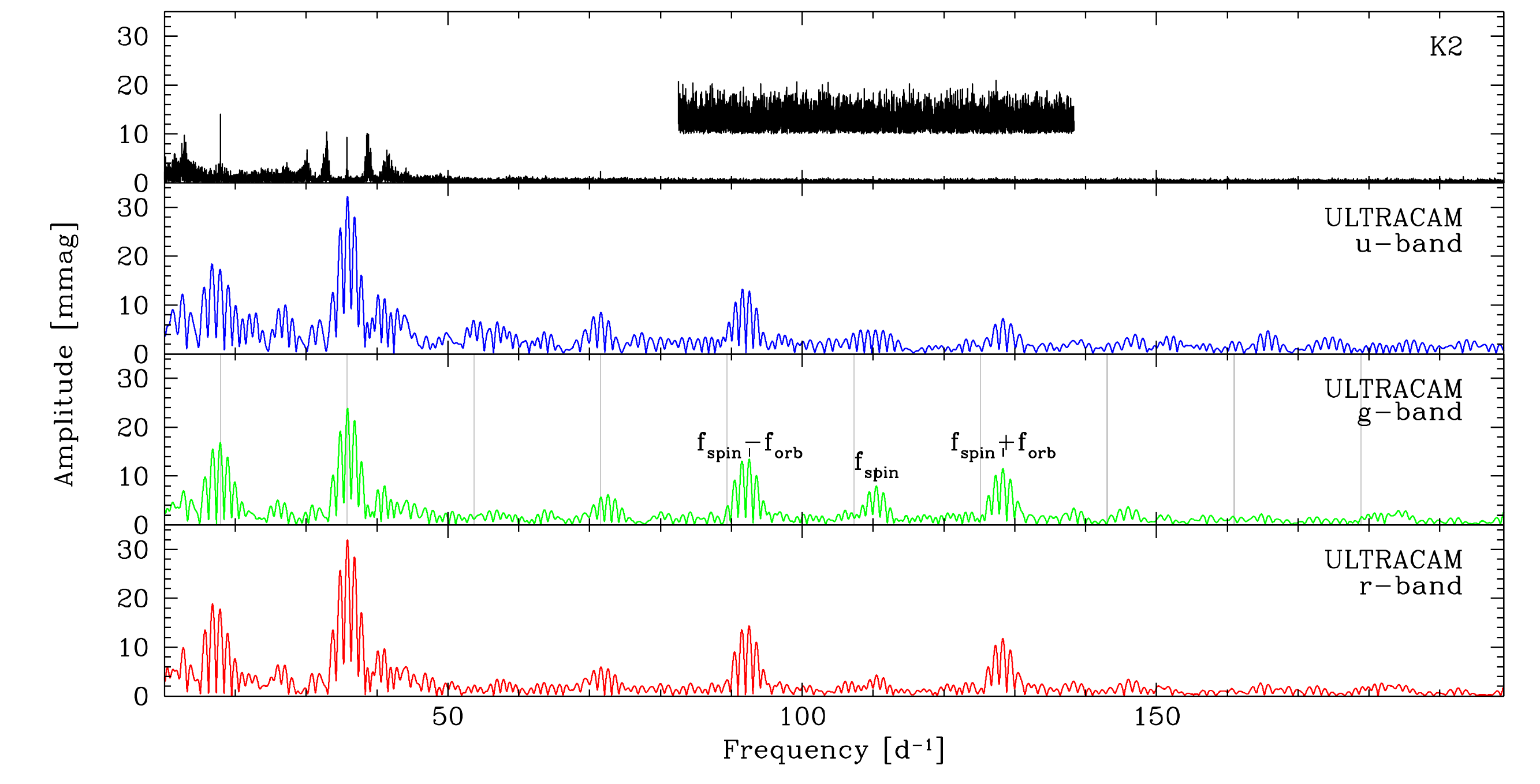}
    \caption{Amplitude spectra computed from the $u$-band (blue), $g$-band (green), and $r$-band (red) ULTRACAM light curves, the amplitude spectrum of the \textit{K2} data is shown for comparison in the top panel. The fine splitting in the ULTRACAM amplitude spectra arises from the aliases introduced by the gap between the two individual observing sequences, harmonics of the orbital frequency are indicated by gray vertical lines in the ULTRACAM $g$-band amplitude spectrum. Additional signals detected in the ULTRACAM data are a coherent period at 13\,min, which we interpret as the white dwarf spin, as well as two sidebands separated by the orbital frequency. The top panel shows a zoom into the \textit{K2} amplitude spectrum, multiplied by ten and offset by ten units, emphasising the absence of the spin and sideband signals during the \textit{K2} observations.}
    \label{fig:power_ugr_k2_wide}
\end{figure*}

\begin{figure*}
\includegraphics[width=0.8\textwidth]{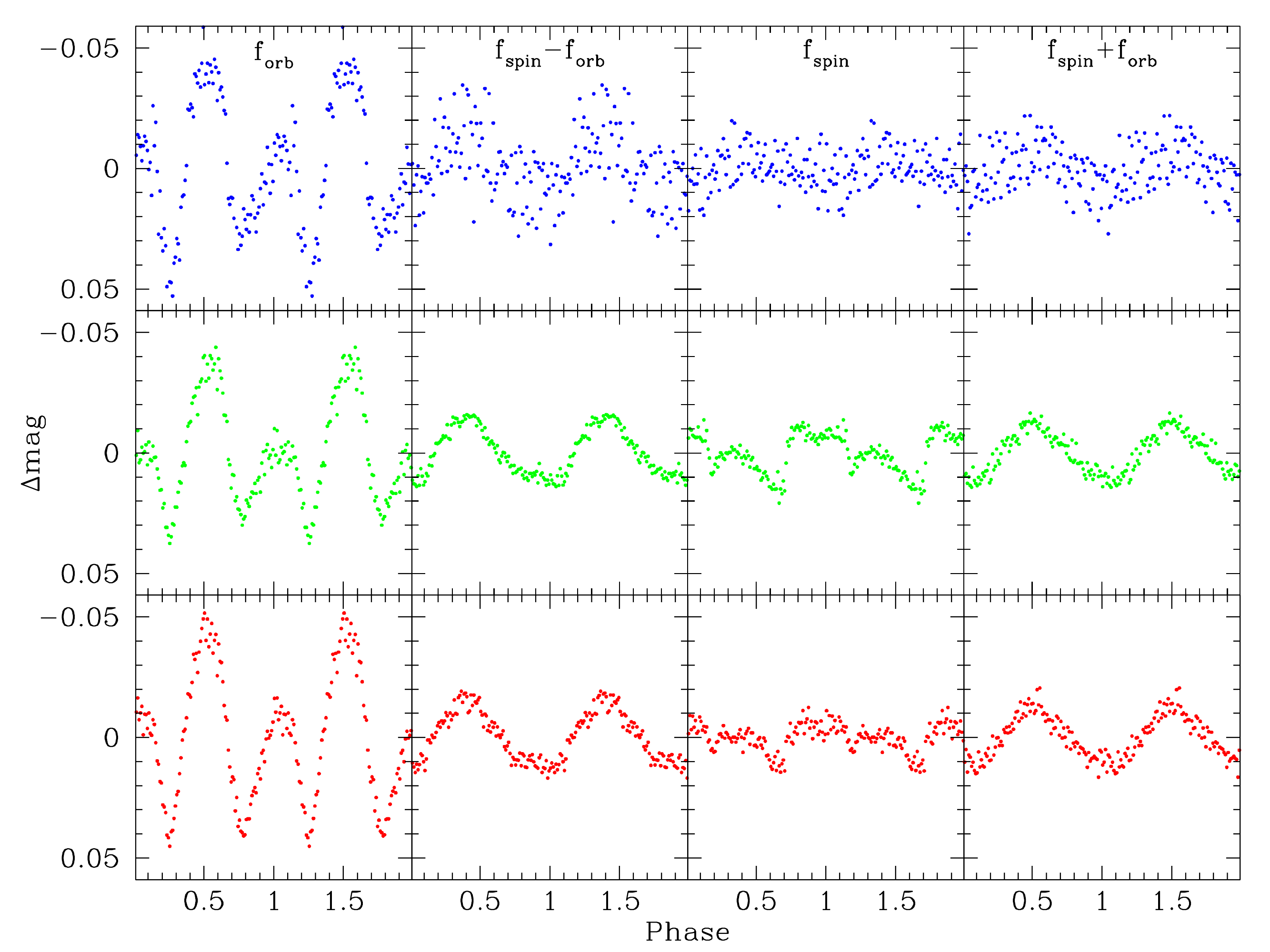}
\caption{The ULTRACAM $u$-band (blue), $g$-band (green), and $r$-band (red) light curves phase-folded on the orbital frequency (left), on the white dwarf spin frequency (second-from-right) and the two side-bands, $\mathrm{f_{spin}-f_{orb}}$ (second-from-left) and $\mathrm{f_{spin}+f_{orb}}$ (right). Prior to the phase-folding on each of these individual four frequencies, the ULTRACAM light curves were pre-whitened with the three remaining frequencies.}
\label{fig:fold_ugr}
\end{figure*}

\subsection{Superoutburst and superhumps\label{subsec:superoutburst}}
The superoutburst of SDSS1238 lasted about 15 days (Figure~\ref{fig:superoutburst_light curve}), although the decline to quiescence could not be followed because of the decreasing visibility of SDSS1238. Because of adverse weather conditions, the photometric observations have a gap of about three days. Superhumps were detected beginning on August 5 (JD $2\,457\,970.5$), but their amplitudes were greatly reduced after the gap in the observations. 

We analysed the superoutburst light curve using the \textsc{MIDAS/TSA} package implemented by \citet{schwarzenberg-czernyetal96-1}. A first cursory analysis showed that the superhump period  ($P_\mathrm{SH}$) was evolving throughout the outburst, and that due to the lower amplitude of the superhumps after the three-day gap, the quality of the data was insufficient to derive reliable results for these final six observing runs. We therefore restricted the following analysis to the observations obtained in the first $\simeq5.5$\,d of the outburst, computing discrete Fourier transforms for groups of three consecutive observing runs, typically spanning $\sim0.25-0.5$\,d, stepping through the entire data set. For each group of observations, we determined the period of the strongest signal in the power spectrum, and subsequently fitted a sine wave to the data to determine an uncertainty on the period. The resulting superhump periods and their amplitudes are shown in the inset panels in Figure~\ref{fig:superoutburst_light curve}. Given the overlap between the grouping of the observing runs, adjacent measurements are not statistically independent. 

Following the definition of stage A, B and C of superhump evolution from \citet{Kato_et_al_2009}, we concluded that stage A, during which the superhump period is the longest, was missed and that the observations were carried out during the middle/late stage B, during which $P_\mathrm{SH}$ is growing and the superhump amplitude is decreasing (see insets in Figure~\ref{fig:superoutburst_light curve}). We adopted the average superhump period throughout this phase, $P_\mathrm{SH} = 81.987 \pm 0.005\,\mathrm{min}$.  Assuming this value and an orbital period of $\porb = 80.5200 \pm 0.0012\,$min (see Section~\ref{subsec:k2_analysis} and \citealt{Zharikov_2006}), we used equation~10 and equation~5 from \citet{Kato_Osaki_2013} for mean stage B superhumps to estimate the system mass ratio, which results in $q = 0.08 \pm 0.01$. We note that the side-bands to the orbital period detected in the $K2$ light curve (Section\,\ref{subec:k2_obs}), interpreted as evidence for a persisting eccentricity of the disc during the quiescent phase, comfortably fall within the period range of the early superhumps detected during the superoutburst.

\subsection{The return to quiescence}\label{subsec:ultracam_analysis}
The ULTRACAM light curves (Figure~\ref{fig:ultracam_lc}) show a more complex structure compared to the pre-outburst $K2$ light curve (Figure~\ref{fig:k2_lc}), with a variability on time scales of minutes and no signs of the brightenings.
However, the non-detection of the brightenings could be due to the limited duration of the ULTRACAM observations. In order to verify whether this is  case, we randomly sampled the $K2$ light curve, selecting chunks of data with a time window matching that of the ULTRACAM observations (two nights of 7.66 and 6.41 hours each, spaced out by 15 hours).
For each of these chunks, we compared the best-fit model to the $K2$ light curve defined in Section~\ref{sec:brightening} with the maximum variation observed in the $g$ band (i.e. the filter with the closest profile to that of the $K_p$ filter) with ULTRACAM in April 2018, i.e. $\simeq 0.2\,$mag. We then assumed that a brightening would have been detected by ULTRACAM either in the first, or the second night, or in both, when the variation of the best fit model in the given temporal window is greater than $\simeq 0.2\,$mag. 
We repeated this procedure 10\,000 times and found that in $\simeq 95$ per cent of the cases a brightening was detected in both nights, in $\simeq 5$ per cent of the cases the brightening was detected either on the first or the second night and in only $\simeq 0.001$ of the cases no brightening was detected. 
This demonstrates that brightenings occurring with the same recurrence time and amplitude as those observed in the $K2$ data would have been detected during the ULTRACAM observations. However, we cannot exclude that after the superoutburst the brightenings could be occurring on longer time scales. Additional observations with longer time span are needed to investigate this possibility.

As for the \textit{K2} and superoutburst light curves, we used the MIDAS/TSA package of \citet{schwarzenberg-czernyetal96-1} to compute discrete Fourier transforms of the ULTRACAM light curves. The resulting amplitude spectra (Figure\,\ref{fig:power_ugr_k2_wide}) contain again a strong signal at half the orbital period, 40.26\,min, and the corresponding higher harmonics, as well as a set of three coherent signals that were not detected in the \textit{K2} data, with periods of $11.221\pm0.001$\,min,  $13.038\pm0.002$\,min, and $15.561\pm0.002$\,min (where the error was determined from fitting a sine functions to the ULTRACAM data). These signals are clearly detected in both nights and are unambiguously offset from higher harmonics of the orbital period. The frequency splitting between the three signals is, within the uncertainties, equal to the orbital frequency. Phase-folding the 13.038\,min modulation results in a highly asymmetric light curve (Figure\,\ref{fig:fold_ugr}, middle panels), whereas the two other signals result in quasi-sinusoidal phase-folded light curves (Figure\,\ref{fig:fold_ugr}, left and right panels). Also noticeable are the colour dependence of the three signals. The 13.038\,min modulation has the largest amplitude in the $g$-band, the amplitude of the 11.221\,min modulation shows no strong variation with colour, and the 15.561\,min modulation has a lower amplitude in the $u$-band. The detection of a coherent signal suggests that we see the white dwarf spin, and we interpret the central signal, 13.038\,min as the spin, and the two side-bands related to either re-processing of light emitted from the white dwarf within the system, or linked to modulations in the accretion flow through the spiral arms (see Section~\ref{subsec:discussion_double-humps}).
At first sight, the asymmetric shape of the 13.038\,min modulation is reminiscent of the light curves of some magnetic CVs, where the sharp changes in flux are due to cyclotron beaming (e.g. SDSS\,085414.02+3905037.2 in figure\,1 of \citealt{Dillon_et_al_2008}). However, considering the upper limit on the magnetic field strength in SDSS1238, $B<100$\,kG, places the cyclotron fundamental in the mm wavelength range, ruling out that the optical modulation detected in the ULTRACAM light curves results from cyclotron beaming of a higher harmonic.

\subsection{System parameters and optical spectral fitting}\label{subsec:system_par}
\begin{figure}
  \includegraphics[width=0.48\textwidth]{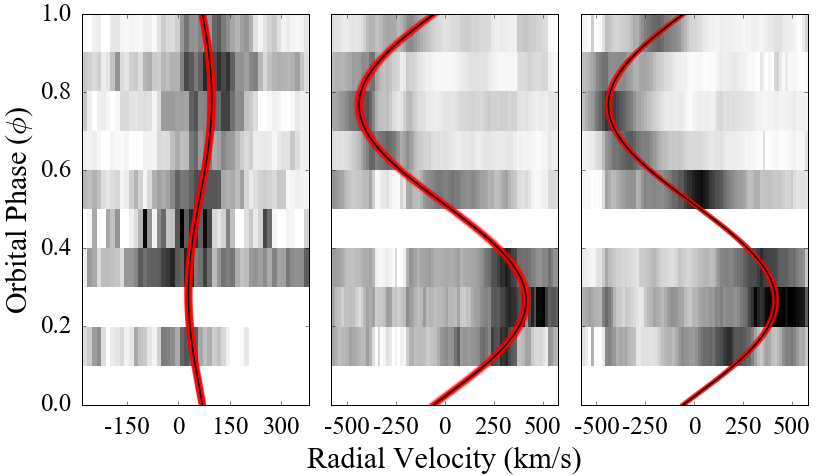}
 \caption{Trailed spectra for the \ion{Mg}{ii} line ($4481$ \AA, left) and the \ion{K}{i} lines ($12\,432/12\,522$ \AA, middle and right). Overplotted are the best fit models (black) along with their uncertainties (red).}\label{fig:trailed}  
\end{figure}
We identified the \ion{Mg}{ii} absorption line at 4481 \AA\ in the X--shooter spectra, which originates in the white dwarf photosphere, and several absorption features arising from the secondary photosphere, including \ion{Na}{i} ($11\,381/11\,403$ \AA), \ion{K}{i} ($11\,690/11\,769$ \AA\, and $12\,432/12\,522$ \AA). Among these, the \ion{K}{i} ($12\,432/12\,522$ \AA) lines are the strongest feature and the only ones visible in all spectra. These lines, together with the \ion{Mg}{ii} line, were used to track the motion of the two stellar components.

Owing to the relatively low SNR ratio of our data ($\simeq 10-20$ and $\simeq 5$ in the individual UVB and NIR spectra) and to the presence of strong residuals from the sky line subtraction in the NIR, we decided to simultaneously fit all the spectra to achieve more robust radial velocity measurements (as done, for example, by \citealt{steven_sdss0857}). 

We first fitted the \ion{K}{i} absorption lines using a combination of a constant and a double Gaussian of fixed separation. We allowed the position of the Gaussians to vary as
\begin{equation}
V = \gamma + K \sin[2\pi(\phi-\phi_0)]
\end{equation}
where $V$ is the radial velocity, $\gamma$ is the systemic velocity, $K$ is the velocity amplitude, $\phi$ is the orbital phase and $\phi_0$ is the zero point of the ephemeris.

From this fit, we determined the $\gamma$ velocity ($\gamma_\mathrm{sec} = -11 \pm 9 \,\kms$) and the velocity amplitude ($K_2 = 428 \pm 12 \,\kms$) of the secondary star (middle and right panel of Figure~\ref{fig:trailed}). Following the same method, we fitted the \ion{Mg}{ii} absorption line using a combination of a constant and a single Gaussian and found the $\gamma$ velocity ($\gamma_\mathrm{WD} = 57 \pm 4 \,\kms$) and the velocity amplitude ($K_1 = 54 \pm 6 \,\kms$) of the white dwarf. However, the low SNR of the data and the presence of the \ion{He}{i} emission line at $4471\,$\AA\, prevent an accurate fit of the \ion{Mg}{ii} line in some of the spectra. Therefore, to better constrain the fit, we combined $K_2$ with the mass ratio determined from the superhump analysis, obtaining the white dwarf velocity amplitude $K_1 = 34 \pm 5\,\mathrm{km s}^{-1}$. We then repeated the fit to the \ion{Mg}{ii} absorption line assuming $K_1$ to be normally distributed around this value, thus obtaining the gamma velocity of the white dwarf, $\gamma_\mathrm{WD} = 63 \pm 4 \,\kms$ (left panel of Figure~\ref{fig:trailed}). The results of this fit are consistent, within the uncertainties, with the values derived by allowing $K_1$ as free parameter in the fit to the \ion{Mg}{ii} line. We assume in the following discussion $\gamma_\mathrm{WD}$ and $K_1$ as derived from the fit assuming the constraint from the mass ratio.
 
\begin{figure*}
 \includegraphics[angle=-90,width=\textwidth]{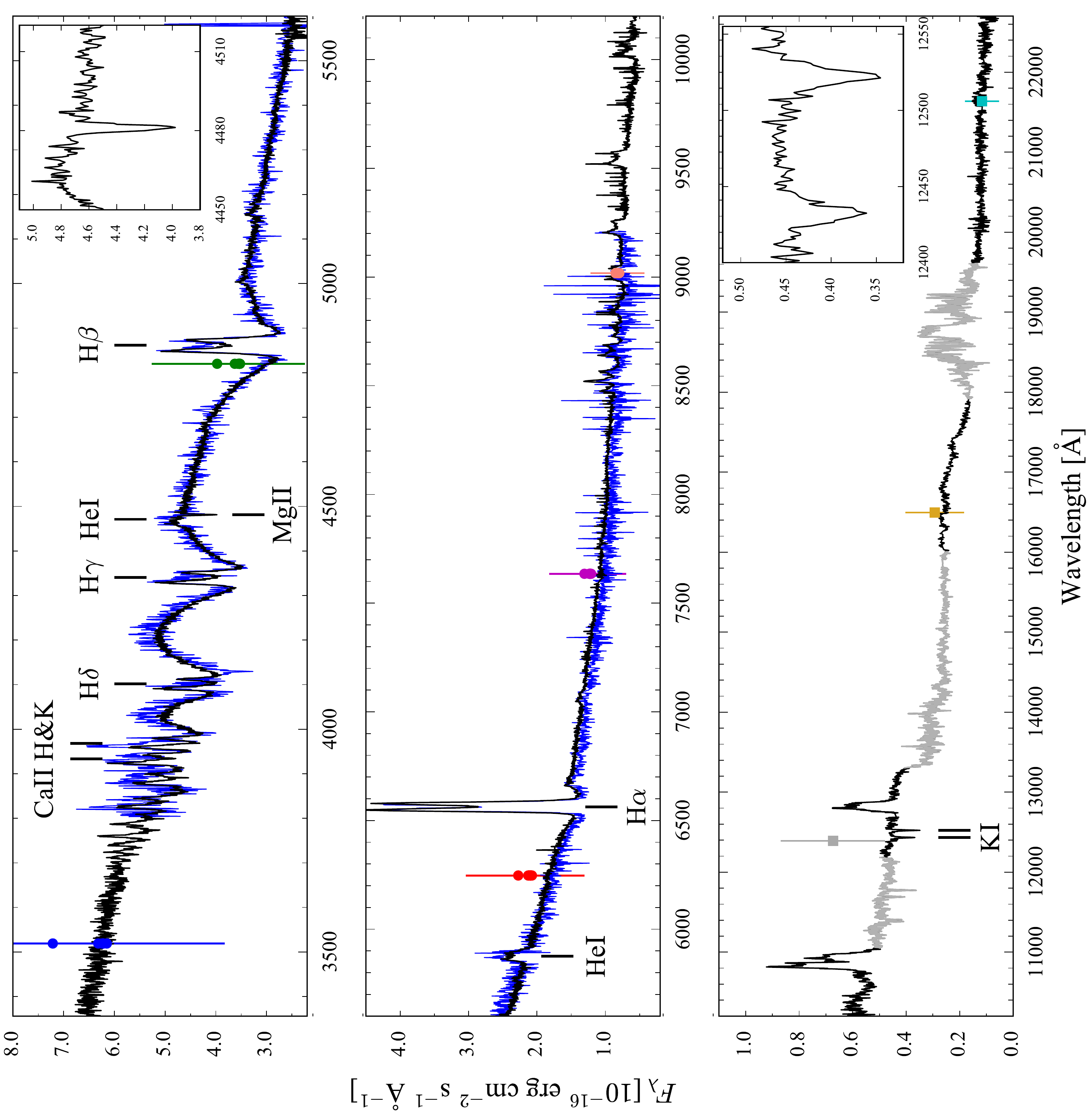}
 \caption{VLT/X--shooter average spectrum (black, upper panel UVB, central panel VIS, bottom panel NIR) of SDSS1238. Regions strongly affected by telluric absorption are drawn in gray. The two insets show the white dwarf \ion{Mg}{ii} ($4481$ \AA) and the secondary \ion{K}{i} ($12\,432/12\,522$ \AA) absorption features used to measure their radial velocities. For comparison, the SDSS spectrum is shown in blue (no offset in flux has been applied). The agreement between the two spectra illustrates the good relative and absolute flux calibration of the X-shooter data. The circles and the squares are the SDSS (blue, green, red, purple and salmon for the $u$, $g$, $i$, $r$ and $z$ band filters) and 2MASS (grey, gold and cyan for the $J$, $H$, $K_s$ band filters) photometry (obtained at different epochs).}\label{fig:sed} 
\end{figure*}

The gamma velocities of the white dwarf and the secondary carry information on the white dwarf mass, since their difference is a measurement of the white dwarf gravitational redshift \citep{Greenstein_1967}:
\begin{equation}
v_\mathrm{grav}(\mathrm{WD}) = \gamma_\mathrm{WD} - \gamma_\mathrm{sec} = 0.635 \, \frac{M_\mathrm{WD}}{M_\odot} \, \frac{\rsun}{\rwd}\,\mathrm{km\,s}^{-1}
\end{equation} 
where $v_\mathrm{grav}(\mathrm{WD})$ and $v_\mathrm{grav}(\mathrm{sec})$ are the gravitational redshift of the white dwarf and the secondary star, respectively. The contribution of the low--mass donors in CVs (typically $M_\mathrm{sec} \lesssim 0.7\,\mathrm{M}_\odot$) to the gravitational redshift in the white dwarf photosphere, $v_\mathrm{grav}(\mathrm{sec}) \simeq 1\, \mathrm{km s^{-1}}$, is negligible compared to the typical uncertainties ($\simeq 10\, \mathrm{km s^{-1}}$) on the $v_\mathrm{grav}(\mathrm{WD})$. Using the zero--temperature mass--radius relationship for white dwarfs with a carbon-oxygen core from \citet{m-r_relationship}, which is justified for the relatively low white dwarf temperature, we found that the mass and the radius of the white dwarf in SDSS1238 are $\mwd = 0.98 \pm 0.06\,\msun$ and $\rwd = 0.0085 \pm 0.0006\,\rsun$, corresponding to $\log(g) = 8.57 \pm 0.06$, where $g$ is expressed in \textit{cgs} units. 
This dynamical measurement of the mass of the white dwarf is consistent with the value derived from the fit to the ultraviolet spectrum \#18, $\mwd \simeq 1\msun$ (Section~\ref{subsec:hst_analysis}) where the contribution from the hot spot on the white dwarf surface is minimum.

\begin{figure*}
 \includegraphics[angle=-90,width=\textwidth]{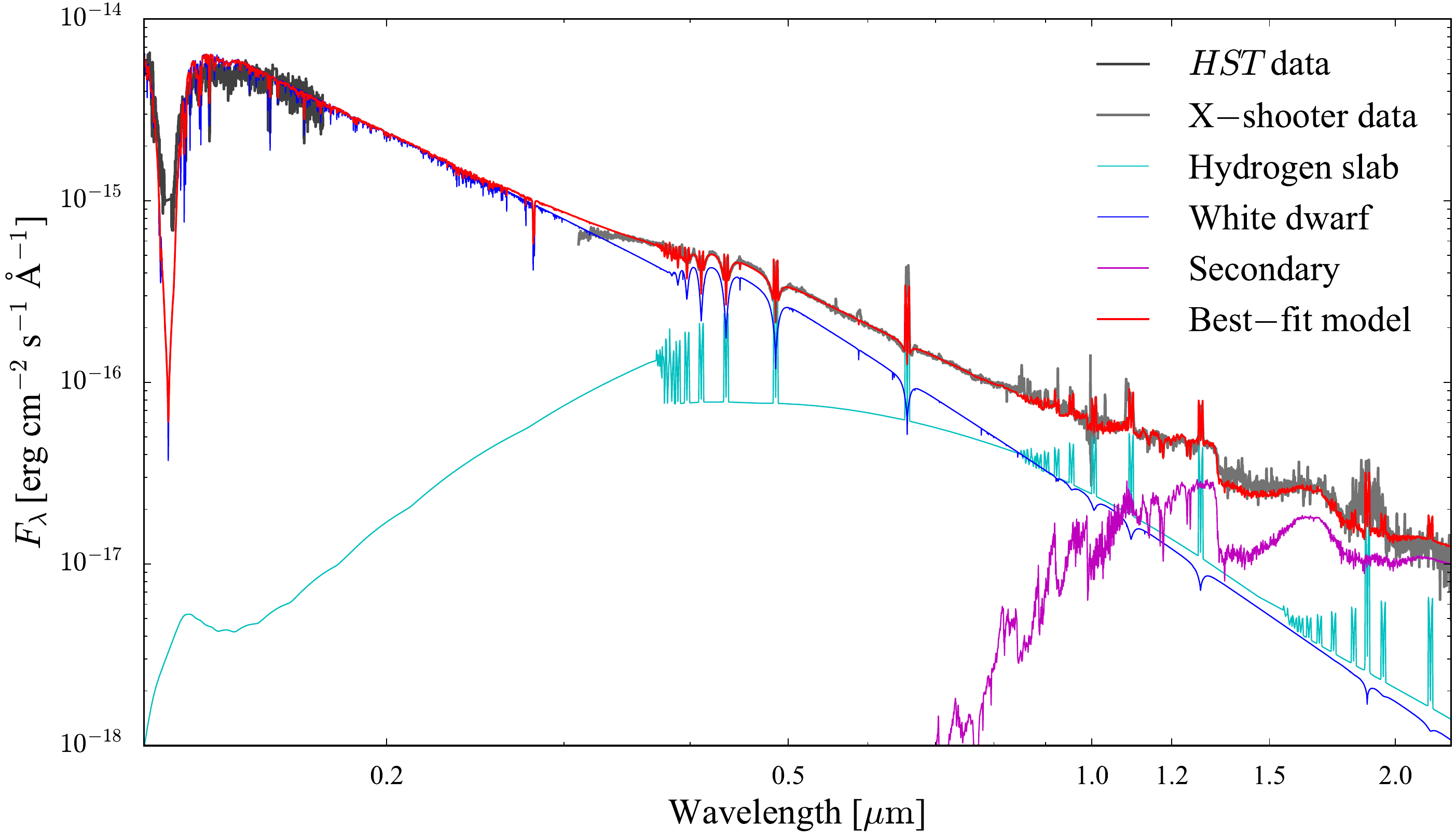}
 \caption{\hst/COS (dark grey) and X--shooter average (light grey, reconstructed as described in the text) spectra of SDSS1238, along with the best--fit model (red, see Table~\ref{table:fit_param} for the relevant parameters), which is composed of the sum of a white dwarf (blue, $\twd = 18\,000\,$K), an isothermal and isobaric pure--hydrogen slab (cyan) and a late--type star (magenta). The \hst data were not included in the fit but are overplotted to show the agreement between the best--fit model and the observed flux level in the ultraviolet (note that the \hst spectrum is the average over the intrinsic variability of SDSS1238, see Section~\ref{subsec:hst_analysis}).}\label{fig:best_fit_sed}
\end{figure*}

\begin{table*}
\caption{Best-fit parameters and their range of variations for the model used to fit the SED of SDSS1238.}\label{table:fit_param}
\begin{center}
\begin{tabular}{lccc}  
\toprule
Parameter                    & Model                                & Range of variation     & Best--fit result      \\
\midrule
$\twd$ (K)                   & \textsc{tlusty} and \textsc{synspec} & $ 10\,000 - 49\,000$   & $18.0(1.4)\times10^{3}$        \\
\midrule
$T_\mathrm{slab}$ (K)        & \citet{Boris_disc, Boris_disc1}      & $5700 - 8000$      & $6.0(3)\times10^{3}$ \\
Pressure (dyn\,cm$^{-2}$)    & \citet{Boris_disc, Boris_disc1}      & $0 - 1000$           & $3.0(8)\times10^{2}$ \\  
Rotational velocity ($\kms$) & \citet{Boris_disc, Boris_disc1}      & $0 - 3000$           & $1.2(7)\times10^{3}$ \\
Geometrical height (cm)      & \citet{Boris_disc, Boris_disc1}      & $10^6 - 10^{12}$         & $7.0(8)\times10^7 $ \\
Slab scaling factor          & \citet{Boris_disc, Boris_disc1}      & $ > 0 $                & $4.7(5)\times10^{-21}$ \\
\midrule
$T_2$ (K)         & \textsc{bt--dusty}                    & $ 1000 - 3600$     & $1.86(14)\times10^{3}$ \\
\bottomrule
\end{tabular}
\end{center}
\end{table*}

The secondary mass can then be calculated from the white dwarf mass and the system mass ratio, $M_2 = 0.08 \pm 0.01 \mathrm{M}_\odot$. From Kepler's third law a measurement of the orbital separation is obtained, $a = 0.63 \pm 0.01\,\rsun$. The presence of strong emission lines suggests that accretion was ongoing at the time of the X--shooter observations and thus the secondary must fill its Roche--lobe. The mass--ratio and the orbital separations yield the Roche--lobe radius which corresponds to the secondary radius \citep{sec_radius}: 
\begin{equation}\label{eq:roche_lobe_radius}
R_2 = \frac{a\,0.49\,q^{2/3}}{0.6\,q^{2/3} + \ln (1\,+\,q^{1/3})} =  0.123 \pm 0.005 \,\rsun
\end{equation}
Finally, while the light curve of SDSS1238 does not show the eclipse of the white dwarf, the presence of double peaked emission lines suggests a relatively high inclination of the system. An estimate of the system inclination $i$ can be derived from the white dwarf mass function: 
\begin{equation}
\frac{(\mwd \sin i)^3}{(\mwd + M_2)^2} = \frac{\porb\,K{^3_2}}{2\pi\,G}
\end{equation}
and results in $i = 55 \pm 3 \degree$.

The night of the X--shooter observations was characterized by the passing of some thick clouds which, combined with the intrinsic variability of SDSS1238 (see Section~\ref{subsec:hst_analysis}), makes it impossible to reconstruct a reliable flux--calibrated average spectrum of the system. To reconstruct the system spectral energy distribution (SED), we selected the three spectra with the highest flux for each arm, corresponding to an orbital phase $\phi \simeq 0.75$ (when the dayside of the secondary is mostly exposed and the irradiated emission is at maximum). We used the radial velocities measured from the \ion{Mg}{ii} line to shift the UVB spectra into the rest frame of the white dwarf and the ones from the \ion{K}{i} lines to shift the NIR spectra into the rest frame of the companion. No absorption features were detected in the VIS spectra, hence no shift was applied to these spectra. We then computed the average spectrum, which agrees well with the accurately flux-calibrated SDSS spectrum and the available photometry of SDSS1238 from 2MASS (Two Micron All--Sky Survey) and SDSS. We scaled the remaining individual spectra to the flux level of this high--flux spectrum and calculated the weighted average, thus obtaining the SED shown in Figure~\ref{fig:sed}. The white dwarf contribution can be recognized from the broad Balmer absorption lines in the UVB arm (upper panel); strong double peaked emission lines reveals the presence of an accretion disc and the donor star contribution can be identified in some absorption features (such as the \ion{K}{i} doublet at $12\,432/12\,522$ \AA) in the NIR portion of the spectrum (bottom panel).

The distance to SDSS1238 $d = 170^{+5}_{-4}\,$pc implies an interstellar reddening of $E(B-V) = 0.01\,$mag (from the three--dimensional map of interstellar dust reddening based on Pan--STARRS\,1 and 2MASS photometry, \citealt{panstar2017}). As shown by \citet{survey_paper}, spectral fits in the ultraviolet are affected by interstellar reddening for $E(B-V) \gtrsim 0.1\,$\,mag, and given that optical observations are less sensitive to interstellar extinction we did not apply any reddening correction.

\begin{table}
\caption{Stellar and binary parameters for SDSS1238.}\label{table:sys_par}
\begin{center}
\begin{tabular}{lc}  
\toprule
System parameter & Value\\ 
\midrule
$\porb$ (min)                           & $80.5200 \pm 0.0012 $ \\
$q$                                          & $0.08 \pm 0.01$       \\
$i$ ($\degree$)                        & $55 \pm 3$            \\
$a$ ($\rsun$)                           & $0.63 \pm 0.01$       \\
$d$ (pc) $^{a}$                                  & $170^{+5}_{-4}$\\
$E(B-V)$ (mag)$^{b}$           & $\gtrsim 0.01$ \\
\midrule
White dwarf parameter & Value\\ 
\midrule
$T$ (K) $^{c}$                          & $\lesssim 14\,600$\\
$M\,(\msun$)                            & $0.98 \pm 0.06$       \\
$R\,(\rsun$)                            & $0.0085 \pm 0.0006$  \\
$\log(g)$                               & $8.57 \pm 0.06$      \\
$v_\mathrm{rot}\,(\kms$)                & $110 \pm 30$          \\
$P_\mathrm{spin}$ (min)  $^{d}$     & $13.038 \pm 0.002$\\
$\langle B \rangle$ (kG)                & $ < 100\,$            \\
$\gamma\,(\kms$)                        & $63 \pm 4$            \\
$K\,(\kms$)                             & $34 \pm 5$            \\
$v_\mathrm{grav}\,(\kms$)               & $74 \pm 10$           \\
\midrule
Secondary parameter & Value\\ 
\midrule
$T$ (K)                                 & $ 1860 \pm 140$         \\
$M\,(\msun$)                            & $ 0.08 \pm 0.01$      \\
$R\,(\rsun$)                            & $ 0.123 \pm 0.005$    \\
$\log(g)$                               & $5.17 \pm 0.07$ \\
$v_\mathrm{rot}\,(\kms$)                & $ 136 \pm 24$         \\
$P_\mathrm{spin}$ (min)                  & $73 \pm 11$           \\
Sp2                                     & L$3 \pm 0.5$          \\
$\gamma\,(\kms$)                        & $-11 \pm 9$           \\
$K\,(\kms$)                             & $428 \pm 12$          \\
\bottomrule
\end{tabular}
\begin{tablenotes}
\item \textbf{Notes.} \textit{(a)} From the \textit{Gaia} parallax $\varpi = 5.9 \pm 0.2\,$mas \citep{Gaia2016,Gaia2018}.  \textit{(b)} From \citep{panstar2017}, for $d = 170\,$pc. \textit{(c)} From \hst/COS spectrum at minimum (see Section~\ref{subsec:hst_analysis}). \textit{(d)} From the ULTRACAM observations (see Section~\ref{subsec:ultracam_analysis}).\\
\end{tablenotes}
\end{center}
\end{table}

Fundamental system parameters, such as the distance and effective temperature, can be measured from a spectral fit to the SED (see e.g. \citealt{sdss1433}). A model to fit the data has to take into account the three light sources in the system, i.e. the white dwarf, the disc and the donor star. We used the grid of white dwarf models described in the Section~\ref{subsec:hst_analysis} and an isothermal and isobaric pure--hydrogen slab model to approximate the disc emission. The model is described in \citet{Boris_disc, Boris_disc1} and is defined by five free parameters: temperature, gas pressure, rotational velocity, inclination and geometrical height. Finally, since the metal content in the white dwarf photosphere may also depend on the accretion rate and diffusion velocity, the white dwarf metallicity only represents a lower limit for the metallicity of the accretion stream, which is stripped from the donor photosphere. Given that no model atmospheres are available for low-mass stars with $Z = 0.5\,\mathrm{Z}_\odot$, we assumed $Z = \mathrm{Z}_\odot$ for the secondary metallicity and we retrieved a grid of \textsc{bt--dusty} \citep{BT-Settle} models for late--type stars from the Theoretical Spectra Web Server\footnote{\href{Theoretical Spectra Web Server}{http://svo2.cab.inta-csic.es/theory/newov/index.php?model=bt-settl}.}, covering $T_\mathrm{eff} = 1000 - 3600\,$K in steps of $100\,$K, for $\log g = 5$ (as derived from the secondary mass and radius in the previous Section) and $Z = \mathrm{Z}_\odot$.

We used a $\chi^2$ minimization routine to fit the data assuming $\mwd~=~0.98~\msun, \rwd~=~0.0085~\rsun, M_2 = 0.08~\msun, R_2~=~0.123\,\rsun$ and $i~=~55\degree$, as determined from the analyses of the radial velocities and the superhump excess, and constraining the white dwarf and the secondary star to be located at the distance inferred from the \textit{Gaia} parallax. Moreover, since the secondary star is irradiated by the hotter white dwarf, its intrinsic flux is only a fraction of the observed one (e.g. \citealt{irradiation, sdss1433}):
\begin{equation}\label{eq:sec_sc}
\frac{F_\mathrm{int}}{F_\mathrm{irr}} = \left( \frac{T_\mathrm{sec}}{T_\mathrm{WD}} \right) ^{4} \left( \frac{a}{R_\mathrm{WD}} \right) ^{2} 
\end{equation}
where $F_\mathrm{int}$ and $F_\mathrm{irr}$ are the intrinsic and the irradiated flux of the secondary, respectively. To take into account the effect of irradiation in our fitting procedure we scaled the secondary model according to this relationship. The free parameters of this fit are:
\begin{itemize}
\item the white dwarf effective temperature. 
\item The secondary star effective temperature.
\item The effective temperature, pressure, rotational velocity, geometrical height and scaling factor of the hydrogen slab. To better constrain the fit, we included as free parameters also the fluxes of the disc emission lines of H$\alpha$, H$\beta$ and H$\gamma$. 
\end{itemize}

The results of this fitting procedure are summarised in Table~\ref{table:fit_param} and the best--fit model is shown in Figure~\ref{fig:best_fit_sed}. The white dwarf effective temperature derived from the SED fit is $\twd = 18\,000 \pm 1400\,\mathrm{K}$, which is in good agreement with the findings by \citet{survey_paper} from the analysis of the average \hst/COS spectrum,  $\twd = 18\,358 \pm 922\,\mathrm{K}$. For the secondary star, we determined an effective temperature of $T_\mathrm{sec} = 1860 \pm 140\,\mathrm{K}$, which implies that the donor is a brown dwarf of spectral type\footnote{We note that this is a spectroscopic definition only, the mass we derive is consistent with being at the borderline between being a very low-mass main sequence star or a brown dwarf. The late spectral type for this mass is consistent with the evolutionary sequences of \cite{Knigge2011} which suggests that the donor stars near the period minimum are under-luminous.} L$3 \pm 0.5$. The best--fit parameters for the hydrogen slab are consistent with the typical values of a quiescent dwarf nova accretion disc \citep{Williams1980,Tylenda1981}.

\begin{figure}
 \includegraphics[width=0.48\textwidth]{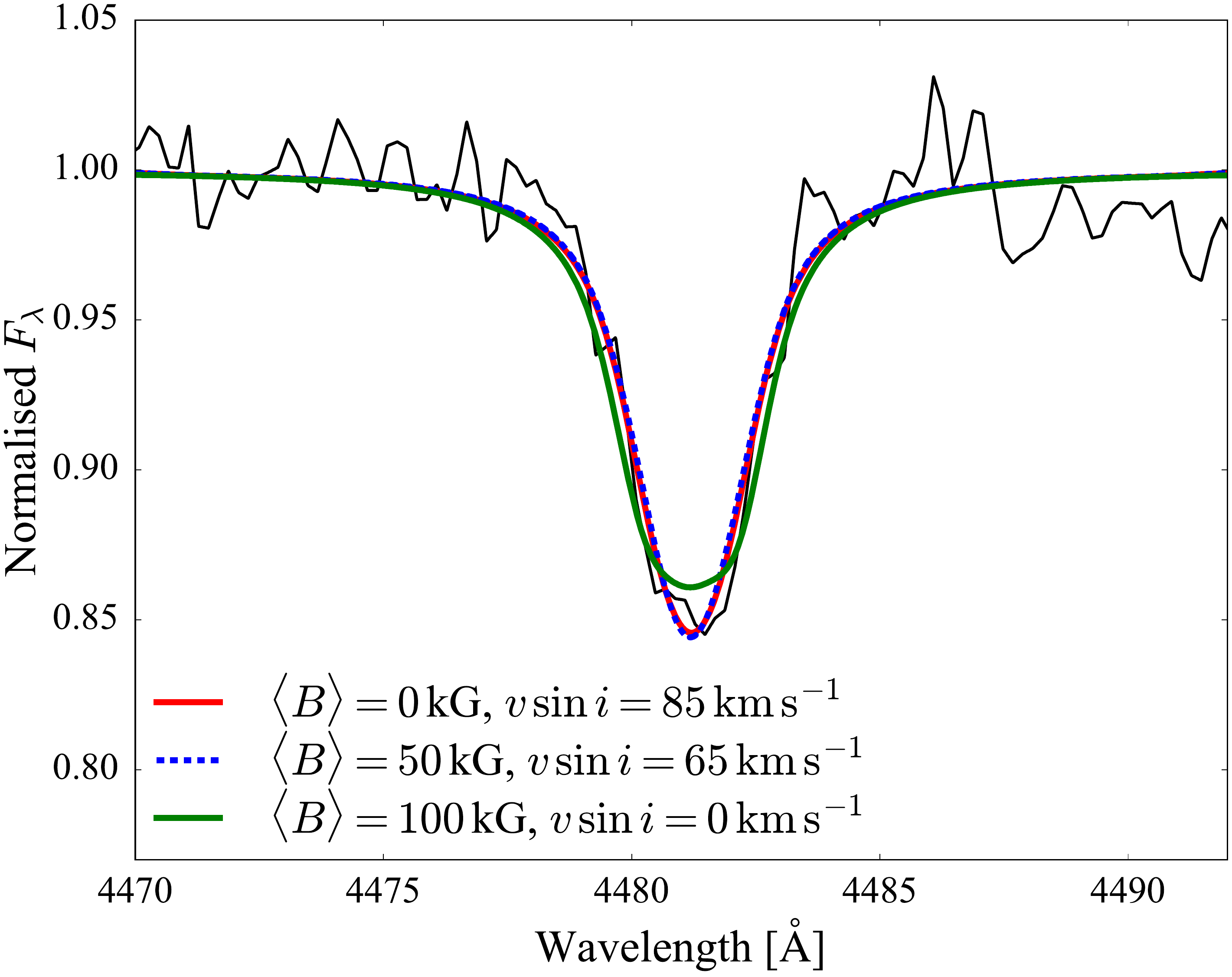}
 \caption{White dwarf \ion{Mg}{ii} absorption line at $4481\,$\AA\, along several atmosphere models computed assuming different strength of the magnetic field and different rotational velocities.}\label{fig:mag_field}
\end{figure}

The scaling factor of the slab provides an estimate of its radius, $R_\mathrm{slab} \simeq 0.2 \mathrm{R}_\odot$. 
Given that the Keplerian velocity of the outer disc corresponds to half of the peak separation of the emission lines, an independent measurement of the disc size comes from the separation between the H$\alpha$ peaks. This results in $R_\mathrm{disc} \simeq 0.2 \mathrm{R}_\odot$, which is consistent with the size of the disc derived from the spectroscopic fit. The size of the Roche lobe of the white dwarf can be determined from Equation~\ref{eq:roche_lobe_radius}, by substituting $q$ with $1/q$, and results in $R_\mathrm{RL} \simeq 0.4 \mathrm{R}_\odot$. 
Our best-fit model for the slab sits inside the Roche lobe of the white dwarf and represents a plausible approximation for the accretion disc in the absence of more complex models. Moreover, the best-fit white dwarf and the slab models contribute, respectively, $\simeq 66$ and $\simeq 34$ per cent to the total emission from the system in the $K_p$ passband filter, which is consistent with what we derived from the $K2$ observations (see Section~\ref{subsec:energetic}). 

The resolution of the X--shooter spectra allows a measurement of the rotational velocity of both stellar components and to put an upper limit on the strength of a possible magnetic field of the white dwarf. In Figure~\ref{fig:mag_field} we show the white dwarf \ion{Mg}{ii} absorption line at $4481\,$\AA, which is broadened by the rotation of the white dwarf and (possibly) by its magnetic field. Assuming $i = 55\,\degree$ and a zero magnetic field, the observed line broadening corresponds to a rotational velocity of $v_\mathrm{rot} = 110 \pm 30\, \kms\,$ (red), and a spin period of $P_\mathrm{spin} = 6 \pm 1$ min. Conversely, assuming zero rotation for the white dwarf and fitting the line with magnetic atmospheric models, the observed line broadening is reproduced by $\langle B \rangle \lesssim 100\,\mathrm{kG}$ (green). Given that the white dwarf is accreting mass and angular momentum, rotation very likely contributes to the broadening of the observed line profile \citep{kingetal91-1} and indeed the spin period of the white dwarf, $P_\mathrm{spin} \simeq 13\,$min (as derived from the analysis of the ULTRACAM data Section~\ref{subsec:ultracam_analysis}), translates into a rotational velocity of $v_\mathrm{rot} \simeq 48\, \kms\,$. This value constrains the magnetic field of the white dwarf in the range $50\,\mathrm{kG} \lesssim \langle B \rangle \lesssim 100\,\mathrm{kG}$ (between the dashed blue and the green line models), while larger magnetic fields imply a broadening due to the Zeeman splitting that is not observed in our data.

Finally, from a Gaussian fit of the \ion{K}{i} lines, we determined the secondary rotational velocity, $v_\mathrm{spin} = 136 \pm 24\,\kms\,$. This yields a spin period of $P_\mathrm{spin} = 73 \pm 11$ min, consistent with the donor star being tidally locked, as expected for a Roche--lobe filling secondary star. 

The system and stellar parameters as derived from the analysis of the X--shooter data are summarised in Table~\ref{table:sys_par}.

\section{Discussion}\label{sec:discussion}
\begin{figure*}\label{fig:doppler}
 \begin{overpic}[width=0.8\textwidth]{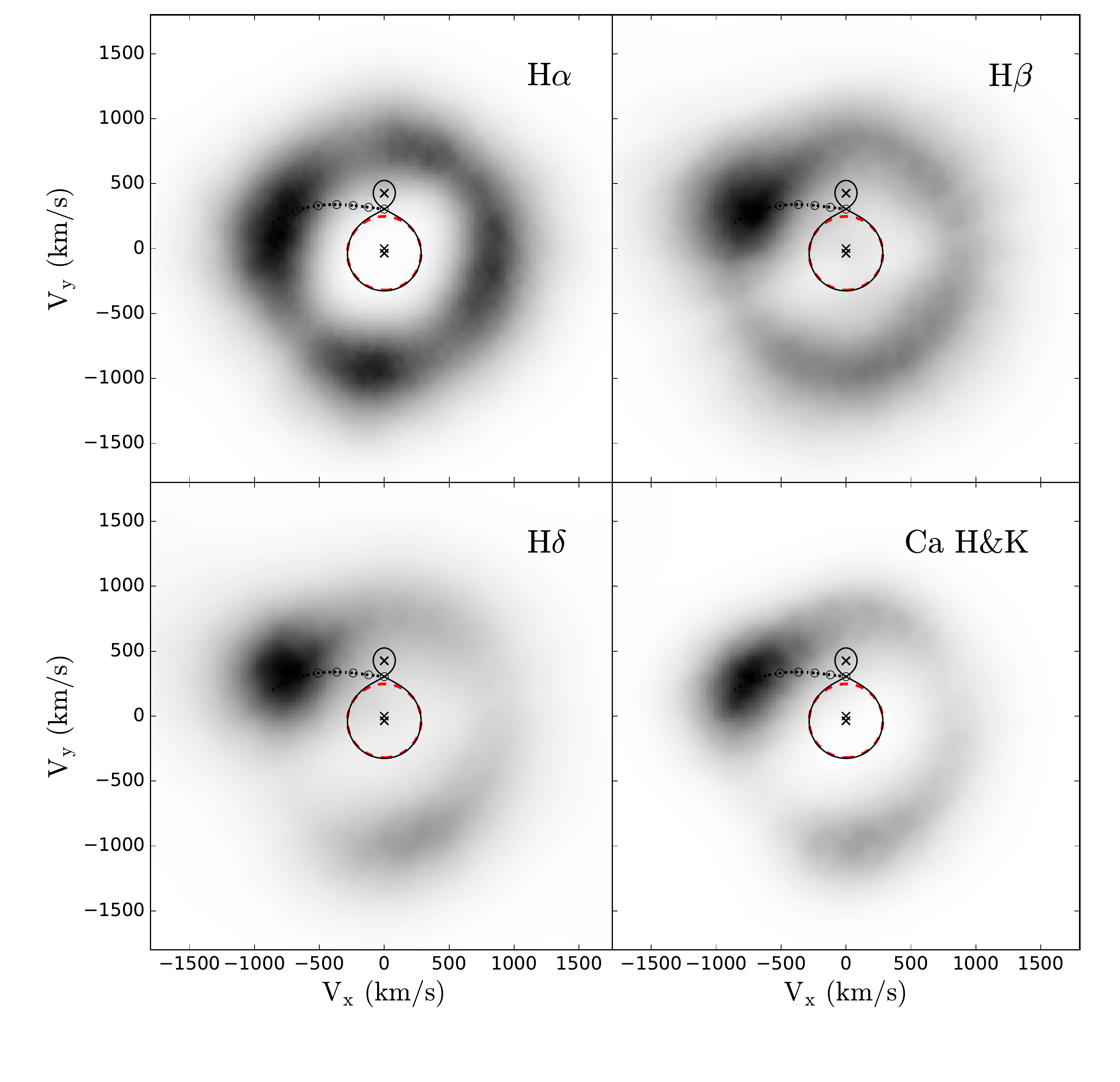} 
 \put (20,77) {\Large \textcolor{red}{1}}
 \put (62,77) {\Large \textcolor{red}{1}}
 \put (63,37) {\Large \textcolor{red}{1}}
 \put (20,37) {\Large \textcolor{red}{1}}  
 \put (33,58) {\Large \textcolor{red}{2}}
 \put (47,72) {\Large \textcolor{red}{3}}
 \put (38,86) {\Large \textcolor{red}{4}} 
 \put (15.5,23.5) {\Large \textcolor{red}{A}}  
 \put (51,39) {\Large \textcolor{red}{B}}   
 \put (57.5,23.5) {\Large \textcolor{red}{A}}  
 \put (93,39) {\Large \textcolor{red}{B}}   
 \linethickness{2pt}
 \put(18,25){\color{red}\vector(4.5,2){6}} 
 \put(50,39.5){\color{red}\vector(-4.5,-2){6}} 
 \put(60,25){\color{red}\vector(4.5,2){6}} 
 \put(92,39.5){\color{red}\vector(-4.5,-2){6}} 
\end{overpic}
\caption{Doppler maps of SDSS1238 computed from the Balmer and the Calcium H\,\&\,K emission lines detected in the X--shooter data. Four regions of enhanced disc emission are clearly visible in the H$\alpha$ map and can be interpreted as the bright spot (1) and the tidal shocks (2-4)  expected in the accretion discs of WZ\,Sge stars \citep{Steeghs_Stehle1999,Kononov_et_al_2015}. The symmetric dips (A and B) detected in the H$\gamma$ and Ca H\,\&\,K lines resemble those detected in the Doppler maps of IP\,Peg (see figure~1 from \citealt{Steeghs_et_al_1997}), suggesting the presence of the spiral density waves. The red dashed line shows the location of the 2:1 resonance radius, located just outside the Roche lobe of the white dwarf.}\label{fig:doppler}  
\end{figure*}
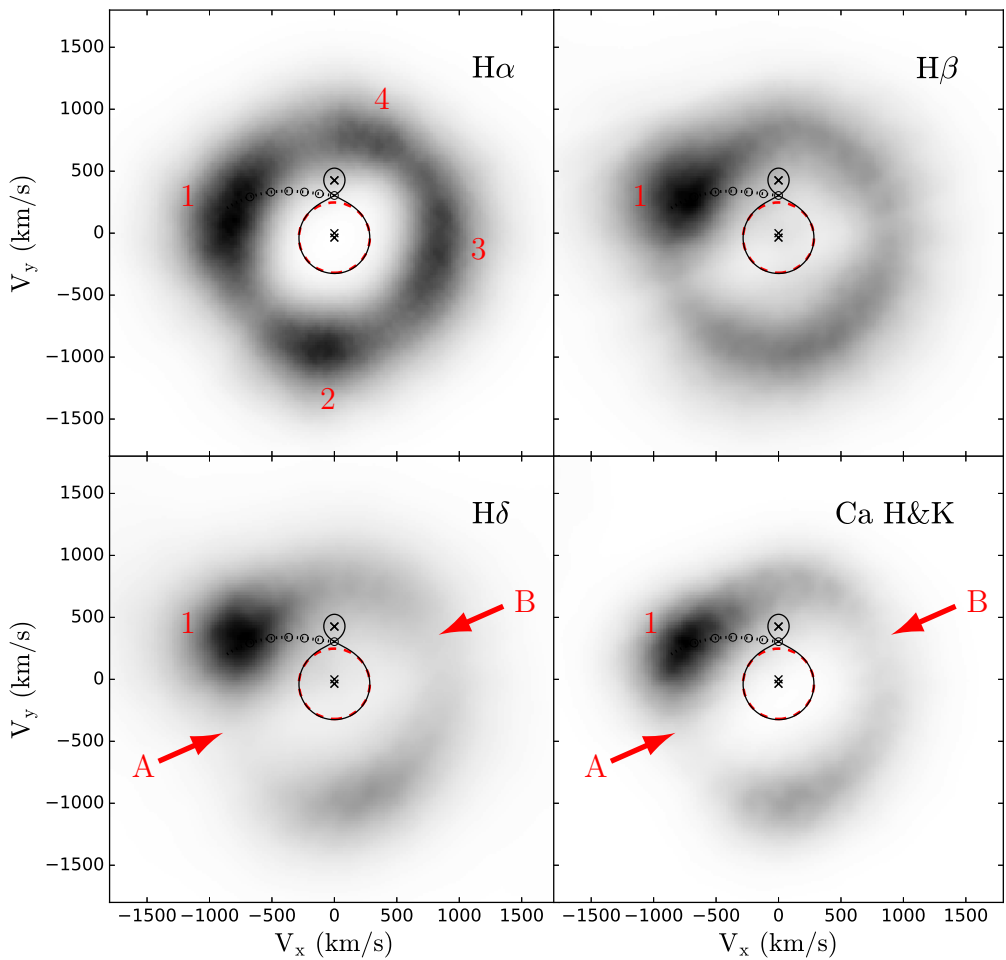

The optical light curve of SDSS1238 shows quasi--sinusoidal double--humps at the orbital period, and quasi--periodic brightenings every $\simeq8.5$\,h, during which the system brightness increases by $\simeq 0.5\,$mag. Moreover, the double--humps grow in amplitude and, as shown from their $O-C$ analysis (Figure~\ref{fig:oc_ampl_short}) their occurrence is delayed when the system brightens, i.e. they undergo phase shifts during the brightenings. We have detected the same behaviour in the ultraviolet data, where the system brightness increases by $\simeq 2\,$mag and $\simeq 1\,$mag during the brightening and the humps, respectively, showing that these photometric modulations originate from the heating and cooling of a fraction of the white dwarf. In the following, we discuss in detail the possible scenarios for the origin of the observed variability of SDSS1238.

\subsection{The double--humps}\label{subsec:discussion_double-humps}
WZ\,Sge stars characterised by low mass accretion rates ($\langle \dot{M} \rangle \simeq 5 \times 10^{-11} \mathrm{M_\odot yr}^{-1}$, \citealt{Dean_and_Boris2009,survey_paper}) and thus have small and optically thin accretion discs. Consequently, also in high inclination systems, the bright spot, i.e. the region of intersection between the ballistic stream and the disc, can be visible along the whole orbital period, even when it is located on the far side, as it can be seen across the optically thin disc \citep{Silber_et_al_1994}. Double--humps could then arise from the variations of the projected area of the bright spot during the orbital motion. However, the \hst\, data clearly demonstrate that both the double--humps and the brightenings originate on the white dwarf, thus ruling out the bright spot scenario.

The modulation of the double-humps at the orbital period implies the presence of two heated regions on the white dwarf that become alternatingly visible to the observer every half orbital period. One or two hot spots on the white dwarf are commonly observed in polars and intermediate polars, CVs hosting magnetic white dwarfs \citep[e.g.][]{stockmanetal94-1, eracleousetal94-1, gaensickeetal95-1, araujo-betancoretal05-2, gaensickeetal06-2}. In these systems, the accreting material couples to the magnetic field lines resulting in localised heating near the white dwarf magnetic poles. As the white dwarf rotates, these spots become periodically visible thus resulting in a flux modulation at the white dwarf spin period. 

However, although the time resolution of the $K2$ data ($\simeq 1\,$min) would easily allow the identification of the white dwarf spin ($\simeq 13\,$min), we do not detect such a signal in the periodogram of the $K2$ light curve (Figure~\ref{fig:tsa}). This implies that, \textit{in quiescence}, the two heated regions on the white dwarf surface must be symmetric with respect to the white dwarf rotation axis.

The quiescent temperature of the white dwarf ($\simeq 14\,000\,$K, Section~\ref{subsec:hst_analysis}) implies that the inner disc of SDSS1238 should be evaporated in a hot corona of ionised gas and, owing to the low mass accretion rate ($\langle \dot{M} \rangle \simeq 4 \times 10^{-11} \mathrm{M_\odot yr}^{-1}$, Section~\ref{subsec:energetic}) onto the white dwarf, the boundary layer is expected to be hot ($\simeq 10^8\,$K), optically thin and to extend longitudinally above the orbital plane \citep{Narayan_Popham_1993}. Our results from the analysis of the $K2$ data are in good agreement with this scenario and provide additional constraints on the geometry of the hot spots on the white dwarf. These must be limited in their longitudinal extension but spread in latitude, in order to also give rise to the observed quasi--sinusoidal modulation during the brightenings when they cover $\simeq 80$ per cent of the visible white dwarf surface (Figure~\ref{fig:fit}, bottom). Such a latitudinal extension is plausible if the inner disc is evaporated into a hot diffuse corona that extends vertically above the disc, as suggested in the coronal siphon flow model by \citet{meyer+meyer-hofmeister94-1}. This hot corona could, in turn, also be the origin of the continuum second component detected in the ultraviolet spectra. Furthermore, given that the white dwarf spin period is much shorter than the orbital period, the cooling time of the hot spots must be significantly shorter than the spin period of the white dwarf, in such a way that the hotter regions are not spread into an heated accretion belt which would not result in the observed photometric humps.
This is possible if the optically thin boundary layer irradiates and advects energy onto/into the white dwarf surface \citep{Abramowicz+1995}, i.e. it behaves like an advection dominated accretion flow \citep{Narayan_Yi_1994,Narayan_Yi_1995}. In this case, the boundary layer is bounded by the optically thick white dwarf surface and, at the interface between them, there is a region that is very hot (i.e. one hot spot)  but with optical depth of the order of unity and therefore can radiate its energy away efficiently and cool down quickly, since the white dwarf layer below it is much denser and colder.
 
In contrast, during an outburst and in the subsequent months, the accretion onto the white dwarf is enhanced, as evidenced by the increased brightness compared to quiescence, $g \simeq 17.6\,$mag, resulting in the hot spots being heated to deeper layers within the white dwarf envelope. These hot spots give rise to the double-humps detected in the ULTRACAM light curve (Figure~\ref{fig:fold_ugr}), just as in the quiescent \textit{K2} observations. However, the deeper heating results in somewhat longer cooling time scales, and azimuthal spreading of the heated regions due to the white dwarf rotation. Owing to the weak magnetic field of the white dwarf, it is possible that these deeper hotter regions assume an asymmetric shape with respect to the rotation axis of the white dwarf, generating a flux modulation at the white dwarf spin period. This would thus explain why the white dwarf spin signal is detected in the ULTRACAM data but not in the $K2$ light curve.

Double--humps during quiescence have been identified in many other short period CVs (see for example \citealt{Patterson_et_al_1996, Augusteijn_Wisotzki_1997, Rogoziecki_Schwarzenberg_2001, Pretorius_et_al_2004, Dillon_et_al_2008}) and are thought to arise from spiral arms in the accretion disc, giving rise to photometric humps locked with the orbital period of the system \citep{Osaki_Meyer_2002}. Spiral arms have been detected in the early phases of the outbursts of IP\,Peg \citep{Steeghs_et_al_1997}, EX\,Dra \citep{Joergens_et_al_2000}, WZ\,Sge \citep{Steeghs_at_al_2001}, and U\,Gem \citep{Groot_2001}, when the enhanced mass transfer rate and angular momentum transfer through the disc allow its expansion up to the 2:1 resonance radius, giving rise to the spiral structure \citep{Lin_Papaloizou_1979}. 
In contrast, during quiescence, the disc is smaller and it is unlikely that it extends out to the 2:1 resonance radius (at least in systems with $q > 0.06$, \citealt{Matthews_et_al_2006}). In this latter case, (at least) two spiral density waves can develop as a consequence of a tidal interaction with the secondary \citep{Spruit_1987,Matsuda_et_al_1990}. The outer regions of the accretion discs are characterised by a Mach number, i.e. the ratio between the orbital velocity and the speed of sound in the gas, of the order of $\simeq 300$ \citep{Ju_et_al_2016}. These high values determine the steepening of the waves into shocks which, as shown by three dimensional magnetohydrodynamic simulations \citep{Ju_et_al_2016,Ju_et_al_2017}, can drive mass accretion onto the white dwarf. Moreover, the Mach number determines the opening angle of the spirals, which results less (or more) wound for low (or high) Mach numbers. 

Spiral density waves can well explain the peculiar light curve of SDSS1238. In this case, two spiral shocks originate in the outer disc and then propagate into the inner region, which is composed of a hot, ionised and optically thin gas and has therefore a low Mach number (of the order of unity). The low viscosity of this region allows the shocks to further propagate inward \citep{Godon_1997} and, since the opening angle of the spirals is determined by the Mach number, they open up and the accreting material hits the white dwarf almost vertically, thus giving rise to two hot spots at the interface between the white dwarf and the inner edge of these spiral shocks.
Each of the two hot spots is visible to the observer once per orbital cycle, resulting in the observed photometric modulation at half the orbital period. The high temperature of the boundary layer allows the spiral arms to be open wound in the inner region of the disc. In the outer disc regions, additional mechanisms could be at work to generate spiral shocks with a larger opening angle than predicted by two dimensional hydrodynamic simulations \citep[e.g.][]{Savonije+1994,Godon_1997}. In fact, recent three dimensional magneto-hydrodynamic simulations have shown that density waves propagate faster than the speed of sound (i.e. at a ``magnetosonic'' speed, given by the quadratic sum of the Alfv{\'e}n speed and the speed of sound). The higher propagation speed has the effect of lowering the Mach number \citep{Ju_et_al_2016,Ju_et_al_2017} allowing the arms to open up.
Moreover, the different amplitudes of the two humps detected in the ULTRACAM light curves eight months after the superoutburst are similar to those observed in the light curve of IP\,Peg in outburst \citep{Harlaftis+1999,Baptista+2000}. This difference arises from the fact that, following the superoutburst, the spiral density waves are asymmetric with respect to the disc centre and generate humps of different strength in the light curve of the system \citep{Harlaftis+1999,Baptista+2000}.

To further investigate the presence of spiral shocks, we computed the Doppler maps from the X--shooter data (Figure~\ref{fig:doppler}). The four regions of enhanced emission detected in the H$\alpha$ Doppler map may be associated with the tidal shocks expected in the accretion discs of low mass ratio CVs \citep{Kononov_et_al_2015}. Moreover, the two symmetric dips observed in the H$\delta$ map resemble those observed on the Doppler maps of IP\,Peg (see figure~1 from \citealt{Steeghs_et_al_1997}), providing further evidence for the presence of spiral density shocks in the accretion disc of SDSS1238. Our Doppler maps are similar to the maps presented by \citeauthor{Aviles2010} (\citeyear{Aviles2010}, see their figure~4), which already suggested the presence of spiral density waves in the accretion disc of SDSS1238. However, their conclusion were based on the assumption of a mass ratio of $q = 0.05$, according to which they argued that the spirals arise from the disc extending up to the 2:1 resonance radius. Thanks to our higher quality data we have been able to measure the mass ratio of SDSS1238, $q = 0.08(1)$. This value implies that the 2:1 resonance radius is located outside the Roche lobe of the white dwarf (red dashed line in Figure~\ref{fig:doppler}) and we therefore concluded that the spirals likely originate from the tidal interaction of the close secondary, as suggested by the sidebands detected at the orbital frequency (Figure~\ref{fig:tsa}), showing that the outer disc maintains a slight eccentricity also during quiescence.

So far, the presence of spiral density shocks in the accretion disc of quiescent CVs has only been determined indirectly \citep[e.g.][]{Yakin_et_al_2010,Bisikalo_2011,Kononov_et_al_2012} and only from the analysis of Doppler maps. Our results represent the first direct evidence for the presence of such shocks during quiescence and their ability to modulate the accretion flow onto the white dwarf. Finally, the requirement of the presence of a hot diffuse corona in the inner disc provides a direct insight into the physics of the boundary layer and an observational support for the models predicting thin boundary layers and the existence of coronal siphon flows.

\subsection{The $8.5\,$h brightenings}\label{brightening_discussion}
During the brightenings, the hot regions become larger as a larger fraction of the white dwarf is heated. These events are not random as, from the analysis of the \textit{K2} light curve, we have identified the presence of a \textit{systematic} mechanism, similar to that driving dwarf nova outbursts.

Accretion discs experience a thermal instability once they undergo partial ionisation of hydrogen, increasing the viscosity and resulting in a temporary increase of the mass flow rate through the disc. This phenomenon is observed in the form of dwarf nova outbursts, during which the white dwarf is heated by the increased accretion rate \citep{Boris1996,sionetal98-1,godonetal04-1,longetal04-1,cooling}. 
In the hot state, the mass flow rate through the disc is higher than the mass loss rate of the secondary, so eventually the column density in the disc drops below a critical rate, resulting in the recombination of hydrogen, a drop in viscosity, and a decrease in the accretion rate onto the white dwarf, which subsequently cools back to its quiescent temperature. Regular dwarf nova outbursts last days to weeks, i.e. much longer than the $\simeq8.5$\,h brightening events in SDSS1238. 

However, a similar phenomenon, on shorter time scales and with smaller amplitudes, could be occurring in SDSS1238. The correlations we identified between the amplitude, the duration and the recurrence time of the brightenings (Figure~\ref{fig:brightnening_properties}) resemble those observed in dwarf nova outbursts \citep{szkody_mattei_1984}, where they originate as a consequence of the piling up of the matter in the accretion disc between each event \citep[e.g][]{Cannizzo_Mattei_1992}. Similarly, in the case of SDSS1238, these could be clues of a disc origin for the brightening events.

Similar thermal instabilities in the accretion disc of SDSS1238 could be related to the presence of the spiral shocks themselves. In order for the shocks to reach the white dwarf, the gas in the disc must have a ratio of specific heats $\gamma < \gamma_\mathrm{max} = 1.45$ \citep{Spruit_1987}. However, if a similar mechanism as that acting in dwarf nova outbursts triggers a thermal instability in the accretion disc, then it is possible that a transition to a state in which $\gamma > \gamma_\mathrm{max}$ occurs. In this configuration, the matter in the disc for which $\gamma > \gamma_\mathrm{max}$ is accreted onto the white dwarf in exactly the same way as the standard disc instability model described before \citep{Spruit_1987}. Therefore the brightening in SDSS1238 could be explained as thermal instabilities in its accretion disc giving rise to a localised increase of the ratio of specific heats of the flowing material. Thermal instabilities are also consistent with the observed correlation between the brightening amplitude and the quiescent magnitude of the system: the more matter is piled up in the disc before triggering the instability, the larger is the amplitude of the brightening (Figure~\ref{fig:amplitude_quiescence}) and, in turn, the longer is the delay before the next instability is triggered (bottom panel of Figure~\ref{fig:brightnening_properties}).

\begin{figure}
\centering
\includegraphics[width=0.7\columnwidth]{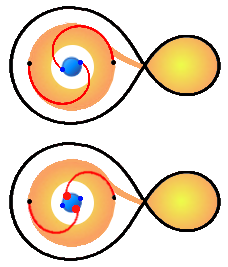} 
  \caption{Schematic drawing of the phase shifts of the double humps during the brightenings. The secondary is filling its Roche lobe and is transferring matter onto the white dwarf via an accretion disc. Owing to tidal instabilities in the accretion disc due to the interaction with the secondary, spiral shocks originating in the outer disc (black dots) propagate towards the centre. In order to generate the humps observed in the light curve during the brightenings, the inner part of the disc must be hot and optically thin, extending in latitude above the orbital plane, illustrated as the white region in the inner part of the disc of this cartoon (the scale is exaggerated). \emph{Top:} during quiescence the spiral arms are phase locked. The corresponding hot spots cover $\simeq 10$ per cent of the white dwarf surface. \emph{Bottom:} during the brightenings, the sound speed increases in the disc and the opening angle of the spiral arms becomes larger. As a result of the enhanced mass accretion rate, the area of the bright spots increases and they undergo phase shifts of up to $\simeq 90$\degree (red dots) with respect to their location in quiescence (blue dots), i.e. they are observed to be delayed with respect to quiescence.}\label{fig:cartoon}
\end{figure}

The combination of thermal instabilities and spiral shocks in the accretion disc might also be the origin of the increase in amplitude of the double--humps and the phase shifts detected during the brightenings in their $O-C$ diagram (Figure~\ref{fig:oc_ampl_short}). A similar increase of the amplitude of the double--humps during the brightnenings has also been observed in SDSS0804 \citep{Szkody2006,Zharikov2008}. During the thermal instability, the temperature, and thus the sound speed, increases. Consequently, the Mach number becomes lower all over the disc and results in a larger opening angle of the spiral arms and the shocks result phase shifted (illustrated in Figure~\ref{fig:cartoon}), up to about $\simeq 90\degree$ at the maximum of the brightening. The spiral density shocks can then modulate the accretion onto the white dwarf through these hot spots \citep{Rafikov_2016,Ju_et_al_2016,Ju_et_al_2017} causing the increase in size of the heated region on the white dwarf during the brightenings.  Our measurement of the increased luminosity during the brightening (Section~\ref{subsec:energetic}) is consistent with an accretion event. After subtracting the quiescence level, the mass accreted on a single average brightening results in $\simeq 4\times 10^{-16}\,\mathrm{M}_\odot$, a plausible quantity of matter that can be accreted from the disc, which usually have masses of the order of $\simeq 10^{-11}\,\mathrm{M}_\odot$ \citep[e.g.][]{Warner}. This indicates that the thermal instability occurs in a small region of the disc and is consistent with the fact that, also during the brightenings, the disc does not outshine the white dwarf, which remains the dominant source of light from the system.

The disappearance of the brightenings in both SDSS1238 (Section~\ref{subsec:ultracam_analysis}) and SDSS0804 after their superoutburst in 2017 (Section~\ref{subsec:superoutburst}) and 2006 \citep{Zharikov2008}, respectively, further reinforces the idea of a disc origin for these events. It is possible that after the superoutbursts, the delicate equilibrium for the brightening to occur has been broken and has not been restored yet.

Finally, thermal instabilities in the accretion disc would also be consistent with the rise of the emission of the second component observed in the \hst spectra at the onset of the brightening (see the third panel of Figure~\ref{fig:fit}), as its emission increases following the enhanced accretion onto the white dwarf.

Alternatively, the brightenings could be related to accretion bursts arising from magnetically gated material. These bursts can occur when the inner part of the disc is truncated by the magnetic field of the white dwarf at a radius $R_\mathrm{i}$ greater than the co--rotation radius $R_\mathrm{c}$. If the white dwarf is rotating faster than the inner regions of the accretion disc, then a centrifugal barrier can inhibit the accretion process. However, if the magnetic field is not strong enough to act as a propeller and eject the material from the system, then the accreting material accumulates outside the centrifugal barrier (see for example figure~4 from \citealt{Scaringi_et_al_2017}). This material exerts a pressure onto the outer edge of the magnetosphere, producing a decrease of $R_\mathrm{i}$. Once $R_\mathrm{i}$ becomes comparable to $R_\mathrm{c}$, a burst of accretion takes place onto the white dwarf \citep{Syunyaev_Shakura_1977,spruit_taam_1993,dangelo_spruit_2012}. 
The onset of this mechanism depends on the relative value of three parameters: the white dwarf spin period, the strength of its superficial magnetic field and the mass accretion rate. In the case of SDSS1238, we found: $P_\mathrm{spin} \simeq 780\,$s, $\langle B \rangle \lesssim 100\,$kG and $\langle \dot{M} \rangle \simeq 4 \times 10^{-11} \mathrm{M_\odot yr}^{-1}$. For these values SDSS1238 falls between the two dashed lines in figure~4 from \citealt{Scaringi_et_al_2017} and thus the models suggest that accretion of magnetically gated material could be occurring in SDSS1238 in a similar way as in MV\,Lyr \citep{Scaringi_et_al_2017}. However, this scenario is challenged by two facts. First, it does not explain the phase shift of the humps during the brightenings. Second, since the magnetically gated material would result in accretion near the magnetic pole(s) of the white dwarf, as we explained before, the white dwarf spin period should be detected in the periodogram of the $K2$ light curve. The non-detection of this signal implies that the white dwarf magnetic field should be aligned with the white dwarf rotation axis, however, this possibility has been ruled out by the ULTRACAM observations.
Moreover, many of the short period CVs are expected to have similar white dwarf spin periods \citep[e.g.][]{Sion1998,mukaietal17-1}, magnetic fields and mass accretion rates to those of SDSS1238 and therefore they should all fall inside the instability zone in figure~4 from \citealt{Scaringi_et_al_2017}. Nonetheless, similar brightenings as those occurring in SDSS1238 have been identified only in three other CVs (SDSS0804, GW\,Lib and GP\,Com), weakening their possible connection with magnetically gated accretion.

Finally, an unlikely but alternative possibility is that the brightenings are caused by a similar mechanism to that causing the flares in isolated pulsating white dwarf \citep{Bell_et_al_2015, Hermes_et_al_2015, Bell_et_al_2016, Bell_et_al_2017}, i.e. changes in the physical condition of the convective layer while the white dwarf cools and crosses the red edge of the instability strip. Although we do not identify pulsations in the \textit{K2} data, it is possible that they have low amplitudes which are below the detection threshold. 

\section{Conclusions}\label{sec:conclusions}
From multi--wavelength observations of SDSS1238, obtained using \hst, the \textit{Kepler/K2} mission, VLT, NTT and the support of the amateur community, we have investigated the peculiar behaviour of the cataclysmic variable SDSS1238.
Its \textit{K2} light curve is characterised by two types of variability: the brightenings and the double--humps. The brightenings occur every $\simeq 8.5\,$h, during which the white dwarf brightness increases by $\simeq 0.5\,$mag for about one hour. The double--humps are a lower amplitude sinusoidal modulation with a period equal to the orbital period. The two modulations are also clearly visible in the ultraviolet light curve obtained from \hst/COS \textsc{time--tag} observations, during which the system brightens by $\simeq 1\,$mag and $\simeq 2\,$mag during the humps and the brightenings, respectively. Moreover, the \hst spectroscopic observations show that both phenomena arise from the presence of two hot regions on the white dwarf surface, which grow in size during the brightenings.

The hot spots on the white dwarf are consistent with the presence of spiral shocks in the accretion disc, which can originate in the outer edge of the disc owing to tidal interaction exerted by the secondary star. Consequently to energy dissipation, the shocks propagate inward and generate two stationary spirals in the accretion disc. Once the shocks reach the inner disc, they open up and hit the white dwarf almost perpendicularly, giving rise to the two hot spots on the white dwarf. During the orbital motion of the systems, these hot regions become alternatingly visible to the observer every half the orbital period, explaining the modulation of the double--humps. Moreover, the $O-C$ diagram of the double--humps shows that, during the brightenings, they undergo phase shifts.

From the analysis of the high-time--resolution photometry of SDSS1238 obtained with $K2$ we identify the presence of a correlation between the amplitude, the duration and the recurrence time of the brightenings which are clues of a disc origin. We hence suggest that the brightenings arise from thermal instabilities in the accretion disc, similar to those commonly observed in dwarf novae in the form of disc outbursts. In the case of SDSS1238, when a thermal instability is triggered in the disc and a fraction of the accreting material becomes ionised, a localised transition of the ratio of specific heats of the flowing material from $\gamma < \gamma_\mathrm{max} = 1.45$ to $\gamma > \gamma_\mathrm{max}$ takes place. This, in turn, causes the rapid accretion of all material with $\gamma > \gamma_\mathrm{max}$ onto the white dwarf and represents the underlying cause for the brightening.  Moreover, the consequent increase of the sound speed results in a larger opening angle of the spirals, causing a shift of the hot spot by $\simeq 90\degree$ with respect to their position on the white dwarf surface during quiescence, thus explaining also the observed phase shift of the double--humps. By modulating the accretion rate onto the white dwarf, the spiral shocks and the thermal instabilities can explain the heating and cooling of the white dwarf observed in the \hst data. 

Our interpretation is further supported by the X--shooter spectroscopy, from which we determine a range of possible value for the magnetic field of the white dwarf, $50\,\mathrm{kG} \lesssim \langle B \rangle \lesssim 100\,\mathrm{kG}$, assuming a spin period of $P_\mathrm{rot} = 13.038 \pm 0.002\,$min (as derived from the ULTRACAM observations). 
This allows us to rule out alternative scenarios, such an intermediate polar in the case of the humps or accretion of magnetically gated material for the brightenings, since these would require the presence of a photometric signal corresponding to the white dwarf spin, which we do not detect in the high--time resolution $K2$ observations. 

Finally, we report the first detected superoutburst of this system, which occurred on 2017 July 31. Superhumps are clearly visible in the light curve of this event, from which we measure the mass ratio, $q = 0.08 \pm 0.01$. This, combined with the phase--resolved X--shooter observations, allows us to determine the physical parameters of the two stellar components. We find that SDSS1238 hosts a massive white dwarf ($M_\mathrm{WD} = 0.98 \pm 0.06\,\msun$) and a late--type companion ($M_2 = 0.08 \pm 0.01\,\msun$) of L$3\pm0.5$ spectral type, which are typical values for a  CV at the period minimum. 

\section*{Acknowledgements}
Based on observations made with the NASA/ESA \textit{Hubble Space Telescope}, obtained at the Space Telescope Science Institute, which is operated by the Association of Universities for Research in Astronomy, Inc., under NASA contract NAS 5--26555. These observations are associated with program GO--12870.

Based on observations made with ESO Telescopes at the La Silla Paranal Observatory under programme under programme ID 095.D-0802(A) and 0101.D-0832(A).

The research leading to these results has received funding from the European Research Council under the European Union's Seventh Framework Programme (FP/2007--2013) / ERC Grant Agreement n. 320964 (WDTracer). 
Support for this work was provided by NASA through Hubble Fellowship grant \#HST-HF2-51357.001-A, awarded by the Space Telescope Science Institute, which is operated by the Association of Universities for Research in Astronomy, Incorporated, under NASA contract NAS5-26555.

This work has made use of data obtained by the PROMPT-8 telescope, owned by National Astronomical Research Institute of Thailand, and operated by the Skynet Robotic Telescope Network.
 
V.S.D., T.R.M. and ULTRACAM are supported by the STFC. M.R.S. acknowledges support from Fondecyt (grant 1181404).
A.A. acknowledges the support of the National Research Council of Thailand (grant number R2560B149).
P.G. wishes to thank William (Bill) P. Blair for his kind hospitality at the Henry Augustus Rowland Department of Physics and Astronomy at the Johns Hopkins University, Baltimore, Maryland, USA. 




\bibliographystyle{mnras}
\bibliography{SDSS1238} 


\bsp	
\label{lastpage}
\end{document}